\documentclass[12pt,preprint]{aastex}
\usepackage{graphicx,amsmath,color,natbib,soul,morefloats,longtable,mathrsfs,amsfonts,amssymb,tabularx,rotating,threeparttable,booktabs,caption,fixltx2e}
\bibpunct{(}{)}{;}{a}{}{,}

\newcommand{\DM}{\mathcal{DM}}
\newcommand{\feh}{{\rm [Fe/H]}} 
\newcommand{\m}{{\it m}}
\newcommand{\co}{{\it c}}
\newcommand{\ldm}{\mathscr{L}\bigl(\mathcal{DM}\bigr)}
\shorttitle{Distance Determinations for the SEGUE K giants}

\shortauthors{Xue et al.}

\begin{document}

\title{The SEGUE K giant survey II: A Catalog of Distance Determinations for the SEGUE K giants in
  the Galactic Halo}

\author{Xiang-Xiang Xue\altaffilmark{1,2}, Zhibo Ma\altaffilmark{3},
  Hans-Walter Rix\altaffilmark{1}, Heather
  L. Morrison\altaffilmark{3}, Paul Harding\altaffilmark{3}, Timothy
  C. Beers\altaffilmark{4}, Inese I. Ivans\altaffilmark{5}, Heather
  R. Jacobson\altaffilmark{6,7}, Jennifer Johnson\altaffilmark{8},
  Young Sun Lee\altaffilmark{9,10}, Sara Lucatello\altaffilmark{11},
  Constance M. Rockosi\altaffilmark{12}, Jennifer
  S. Sobeck\altaffilmark{13}, Brian Yanny\altaffilmark{14}, Gang
  Zhao\altaffilmark{2}, Carlos Allende Prieto\altaffilmark{15,16}}

\altaffiltext{1}{Max-Planck-Institute for Astronomy K\"{o}nigstuhl 17,
  D-69117, Heidelberg, Germany} 

\altaffiltext{2}{Key Lab of Optical Astronomy, National Astronomical
  Observatories, CAS, 20A Datun Road, Chaoyang District, 100012,
  Beijing, China} 

\altaffiltext{3}{Department of Astronomy, Case Western Reserve
  University, Cleveland, OH 44106, USA} 

\altaffiltext{4}{NOAO, Tucson, Arizona, 85719, USA and JINA: Joint Institute for Nuclear Astrophysics}

\altaffiltext{5}{Department of Physics and Astronomy, JFB\#201, The University of Utah, Salt Lake City, 84112, USA}
\altaffiltext{6}{Department of Physics \& Astronomy, Michigan State
  University, East Lansing, MI 48823, USA}

\altaffiltext{7}{Massachusetts Institute of Technology, Kavli
  Institute for Astrophysics and Space Research, 77 Massachusetts
  Avenue, Cambridge, MA 02139, USA} 

\altaffiltext{8}{Department of Astronomy, 4055 McPherson
Laboratory, 140 W. 18th Ave, Columbus, Ohio, 43210, USA and Center for
Cosmology and Astro-Particle Physics, 191 West Woodruff Ave, Columbus,
Ohio 43210, USA}

\altaffiltext{9}{Department of Astronomy, New Mexico State University,
  Las Cruces, NM 88003} 

\altaffiltext{10}{Department of Physics and Astronomy and JINA: Joint
  Institute for Nuclear Astrophysics, Michigan State University,
  E. Lansing, MI 48824, USA} 

\altaffiltext{11}{Osservatorio Astronomico di Padova, Vicolo
  dell’Osservatorio 5, 35122 Padua, Italy} 

\altaffiltext{12}{Lick Observatory/University of California, Santa
  Cruz, CA 95060, USA} 

\altaffiltext{13}{Laboratoire Lagrange (UMR7293), Universite de Nice
  Sophia Antipolis, CNRS, Observatoire de la Cote d’Azur, BP 4229,
  F-06304 Nice Cedex 04, France; JINA: Joint Institute for Nuclear
  Astrophysics and the Department of Astronomy and Astrophysics,
  University of Chicago, 5640 South Ellis Avenue, Chicago, IL 60637,
  USA}

\altaffiltext{14}{Fermi National Accelerator Laboratory, P.O. Box 500,
  Batavia, IL 60510 USA}

\altaffiltext{15}{Instituto de Astrof\'{\i}sica de Canarias, 38205 La
  Laguna, Tenerife, Spain} \altaffiltext{16}{Departamento de
  Astrof\'{\i}sica, Universidad de La Laguna, 38206 La Laguna,
  Tenerife, Spain}

\begin{abstract}

We present an online catalog of distance determinations for $\rm 6036$
K giants, most of which are members of the Milky Way's stellar
halo. Their medium-resolution spectra from SDSS/SEGUE are used to derive metallicities and
rough gravity estimates, along with radial velocities. Distance moduli are derived from a
comparison of each star's apparent magnitude with the absolute
magnitude of empirically calibrated color-luminosity fiducials, at the
observed $(g-r)_0$ color and spectroscopic $\rm \feh$. We employ a
probabilistic approach that makes it straightforward to properly propagate the errors in metallicities, magnitudes, and colors into distance uncertainties. We also fold in {\it prior} information about the giant-branch luminosity function and the different metallicity distributions of the SEGUE K-giant targeting sub-categories. We show that the metallicity prior plays a small role in the distance estimates, but that neglecting the luminosity prior could lead to a systematic distance modulus bias of up to 0.25 mag, compared to the case of using the luminosity prior. We find a median
distance precision of 16\%, with distance estimates most precise for
the least metal-poor stars near the tip of the red-giant branch. The precision and accuracy of our distance estimates are validated with observations of globular and open clusters. The stars in our catalog are up to 125 kpc distant from the Galactic center, with 283 stars
beyond 50 kpc, forming the largest available spectroscopic sample of distant tracers
in the Galactic halo.

\end{abstract}



\section{Introduction}
Giants of spectral type K have long been used to map the Milky Way's stellar halo
\citep{Bond1980,Ratnatunga1985,Morrison1990,Morrison2000,Starkenburg2009}. In
contrast to blue horizontal-branch (BHB) and RR Lyrae stars, giant
stars are found in predictable numbers in old populations of all
metallicities; and at the low metallicities expected for the Milky Way's halo
they are predominantly K giants. At the same time, their high
luminosities ($\rm M_{\it r} \sim 1$ to $\rm -3$) make it feasible to
study them with current wide-field spectroscopic surveys to distances
of $\rm > 100$ kpc \citep{Battaglia2007}. The Sloan Extension for
Galactic Understanding and Exploration \citep[SEGUE:][]{Yanny2009b},
which now has been extended to include SEGUE-2 (Rockosi et al. in prep.) specifically targeted K giants for spectroscopy as part of the effort to explore the outer halo of the Galaxy. For simplicity, henceforth we refer to SEGUE and SEGUE-2 collectively as simply SEGUE. The SEGUE data products include sky positions, radial
velocities, apparent magnitudes, and atmospheric parameters
(metallicities, effective temperatures, and surface gravities), but no
preferred distances.

Distance estimates to kinematic tracers, such as the K giants, are
indispensable for studies of Milky Way halo dynamics, such as
estimates of the halo mass \citep{Battaglia2005,Xue2008}, for
exploring the formation of our Milky Way \citep[e.g., probing
  velocity-position correlations,][]{Starkenburg2009,Xue2011}, and for
deriving the metallicity profile of the Milky Way's stellar halo. All of
these studies require not only unbiased distance estimates, but also a
good understanding of the distance errors. However, unlike `standard
candles' (i.e., BHB and RR Lyrae stars), the intrinsic luminosities
of K giants vary by two orders of magnitude, with color and luminosity
depending on stellar age and metallicity.

The most immediate approach to estimating a distance to a K giant with
color {\it c} and metallicity $\rm \feh$ (e.g., from SDSS/SEGUE) is
to simply look up its expected absolute magnitude $\rm M$ in a set of
observed cluster fiducials, $\rm M\bigl({\it c},\feh\bigr)$. This
approach was used, for example, by \citet{Ratnatunga1985},
\citet{Norris1985} and \citet{Beers2000}. Comparison with the apparent
magnitude then yields the distance modulus (denoted by $\rm \DM$) and
distance. In practice, this simple approach has two potential problems. First,
care is required to propagate the errors in metallicities, magnitudes,
and colors properly into distance uncertainties. Secondly, such an
approach does not immediately incorporate external prior information,
such as the luminosity function along the red giant-branch (RGB) and
the overall metallicity distribution of the stellar population
under consideration. Because the luminosity function along the RGB is
steep, and there are a larger number of faint stars rather than bright
stars \citep[$\rm n\bigl(L\bigr)\sim
  L^{-1.8}$;][]{sandquist1996,sandquist1999}, an estimate of the
absolute magnitude, $\rm M(\co, \feh)$, is more likely to produce an
overestimate of the luminosity, and therefore an overestimate of the
distance. Analogously, there are few extremely metal-poor (say, $\rm \feh<-3.0$) or
  comparatively metal-rich (say, $\rm \feh>-1.0$) stars observed in the halo, which implies that a very low
  estimate of $\rm \feh$ is more likely to arise from an underestimate of the
  metallicity of an (intrinsically) less metal-poor
  star\footnote{We use the term 'less metal-poor' for the
    most metal-rich stars within our sample, because even those stars
    have metallicities of only $\rm \feh \sim~-1$, far below the
    average of all giants in our Galaxy.}. As a result, the estimated absolute magnitude will lead to an overestimate the luminosity, and thus an overestimate of the distance. Therefore, in order to exploit K giants such as those from SDSS/SEGUE for various dynamical analyses,
an optimal way to determine an unbiased distance probability
distribution for each sample star is crucial.

A general probabilistic framework to make inferences about parameters
of interest (e.g., distance moduli) in light of direct observational
data and broader prior information is well-established. It has been
applied in a wide variety of circumstances, and recently also applied
to the distance determinations for stars, including giant stars in the
RAVE survey \citep{burnett2010,burnett2011}. \citet{burnett2010}
described how probability distributions for all the `intrinsic'
parameters (e.g., true initial metallicity, age, initial mass,
distance, and sky position) can be inferred using Bayes'
theorem, drawing on the star's observables and associated errors
thereon. Here we focus on a somewhat more restricted problem: the
distances to stars on the red giant branch (RGB), which we can presume to be `old' ($\rm
>5$ Gyr). Like any Bayesian approach, our approach is optimal in the
sense that it aims to account for all pertinent information, can
straightforwardly propagate the errors of the observables to distance
uncertainties, and should avoid systematic biases in distance
estimates. This approach also provides a natural framework to account
for the fact that distance estimates will be less precise for stars
that fall onto a `steep' part of the color-magnitude fiducial,
such as metal-poor stars on the lower portion of the RGB.

The goal of this paper is to outline and implement such a Bayesian
approach for estimating the best unbiased probability distribution
of the distance moduli $\rm \DM$, for each star in a sample of $\rm
6036$ K giants from SDSS/SEGUE. This distribution can then be
characterized by the most probable distance modulus, $\rm \DM_{peak}$,
and the central 68\% interval, $\rm \Delta \DM$. At the same time,
this approach also yields estimates for the absolute
magnitude $\rm M$, heliocentric distance $d$, Galactocentric distance
$\rm r_{GC}$, and their corresponding errors.

In the next section, we introduce the selection of the SEGUE K giants
and their observables. In \S 3, we describe a
straightforward (Bayesian) method to determine the distances. The
results and tests are presented in \S 4. Finally, \S 5 presents our
conclusions and a summary of the results.


\section{Data}

SDSS and its extensions use a dedicated 2.5m telescope
\citep{Gunn2006} to obtain $ugriz$ imaging \citep{Fukugita1996, Gunn1998, York2000,
  Stoughton2002, Pier2003, Eisenstein2011} and resolution (defined as
$\rm R = \lambda/\Delta \lambda$) $\sim$2000 spectra for 640 (SDSS spectrograph) or
1000 \citep[BOSS spectrograph;][]{Smee2013} objects over a 7 square degree field. SEGUE, one of
the key projects executed during SDSS-II and SDSS-III, obtained some 360,000
spectra of stars in the Galaxy, selected to explore the nature of
stellar populations from 0.5 kpc to 100 kpc \citep[][and Rockosi et al. in prep.]{Yanny2009b}. Data from SEGUE is a significant
part of the ninth SDSS public data release, DR9 \citep{Ahn2012}.

SDSS DR9 delivers estimates of T$\rm _{\rm eff}$, log g, $\rm
\feh$ and [$\alpha$/Fe] from an updated and improved version of the
SEGUE Stellar Parameter Pipeline \citep[SSPP,][]{Lee2008a,Lee2008b,
  AllendePrieto2008, Smolinski2011, Lee2011}. The SSPP processes the
wavelength- and flux-calibrated spectra generated by the standard SDSS
spectroscopic reduction pipeline \citep{Stoughton2002}, obtains
equivalent widths and/or line indices for more than 80 atomic or
molecular absorption lines, and estimates T$\rm _{\rm eff}$, log g,
and $\rm \feh$ through the application of a number of complementary
approaches \citep[see][for detailed description of these techniques
  and Rockosi et al. in prep. for recent changes and
  improvements of the SSPP]{Lee2008a}.

\begin{figure}
\includegraphics[width=\textwidth]{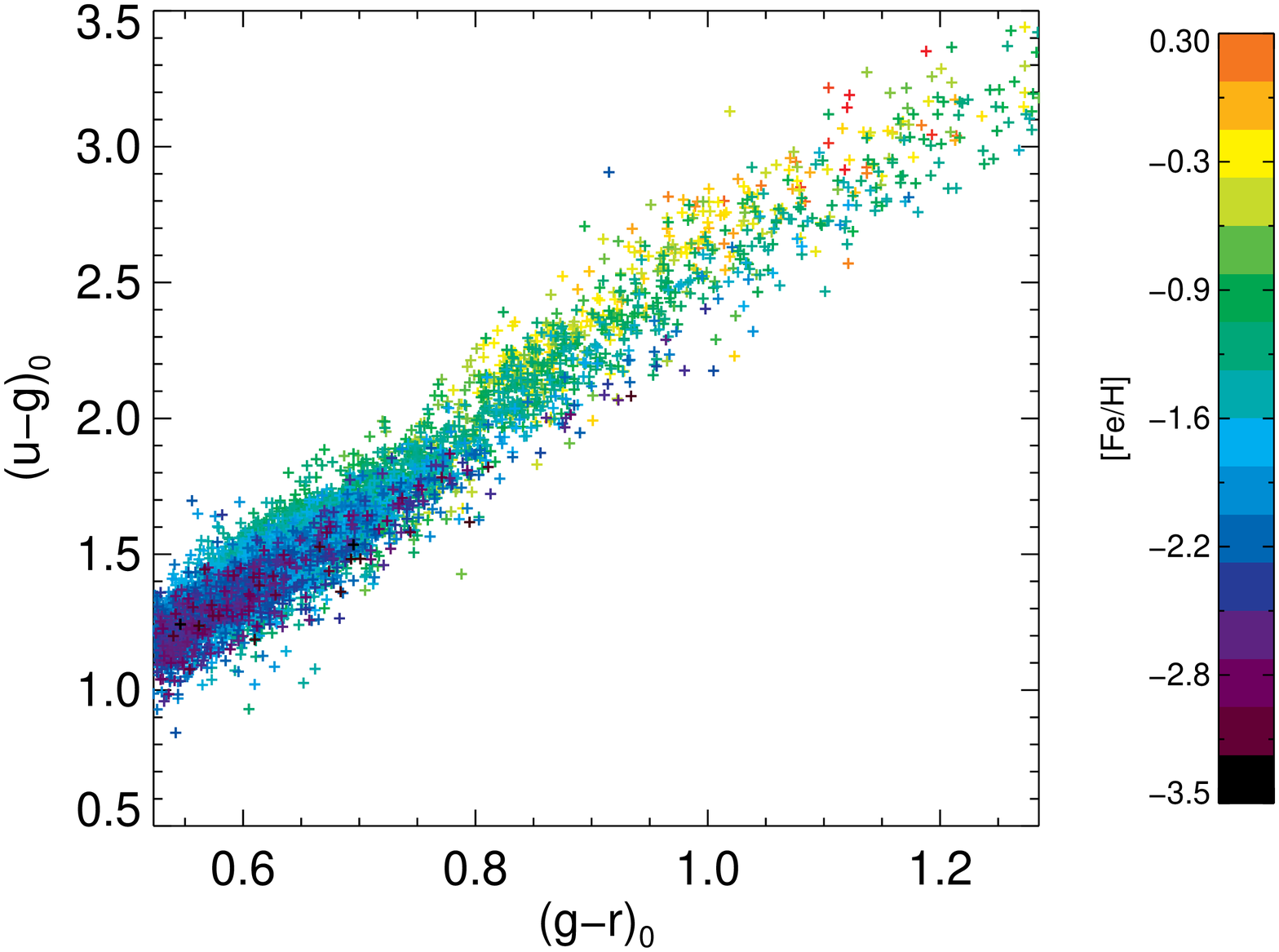}
\caption{Color-color diagram for confirmed K giants in our sample,
  with DR9 SSPP estimates for $\rm \feh$ color-coded as shown in the
  vertical bar.}
\label{f:fcolor}
\end{figure}

The SEGUE project obtained spectra for a large number of different stellar
  types: 18 for SEGUE-1 \citep[see][for details]{Yanny2009b} and 11
  for SEGUE-2 (Rockosi et al. in prep.). Three of these
  target types were specifically designed to detect K giants: these are designated ``l-color K giants'', ``red K giants'', and ``proper-motion K giants''. The K-giant targets from these three categories all have $0.5<(g-r)_0<1.3$, $0.5<(u-g)_0<3.5$ (shown as Firgue~\ref{f:fcolor}), and proper motions smaller than 11 mas/year. Figure 10 of \citet{Yanny2009b} shows the
  regions of the $u-g/g-r$ plane occupied by the three target types:
  each category focuses on a particular region\footnote{Exact criteria
    for each target type can be found at
    \\https://www.sdss3.org/dr9/algorithms/segue\_target\_selection.php/\#SEGUEts1}. In
  brief, the l-color K-giant category uses the metallicity sensitivity
  of the $u-g$ color in the bluer part of the color range to preferentially select
  metal-poor K giants. The two other categories focus on the redder
  stars with $(g-r)_0 > 0.8$: the red K-giant category selects those
  stars whose luminosities place them above the locus of
  foreground stars, while the proper-motion K-giant region is where
  the K giants are found in the locus of foreground stars. In this location, only a
  proper-motion selection can be used to cull the nearby dwarf
  stars, because they have appreciable proper motions compared to the
  distant giants.

We derived the sample of giants presented in this paper as
  follows. Using SDSS DR9 values in all cases, we start by requiring
  that the star has valid spectroscopic measurements of $\rm \feh$ and
  log g from the SSPP. To eliminate main-sequence stars, we make a conservative cut on the SSPP estimate of log g by requiring $\rm \log g < 3.5$. We restrict the star's temperature by requiring that $0.5 < (g-r)_0 < 1.3$. Stars bluer than this color cutoff do not exhibit the
  luminosity-sensitive Mg$b$/MgH feature with sufficient strength to use in the luminosity calibration; the red cutoff delineates the start of the M-star region. We further limit our sample by requiring that the reddening estimate from \citet{Schlegel1998} for each star, E(B-V), is less than 0.25 mag. We also apply additional data-quality criteria for both spectroscopy and photometry, as described in detail in Morrison et al. (in prep.). Most importantly, in addition to the SSPP log g measurement, we calculate a Mg index from the flux-corrected, but not continuum-corrected, SEGUE spectra. This index is a ``pseudo equivalent width", and identical to the Mg index described in \citet{Morrison2003}, except for an adjustment to one of the continuum bands. We compare the value of this index at a given $(g-r)_0$ color with the index values for known globular and open cluster giants of different metallicity to decide whether a star is a giant or a dwarf, taking into account the SSPP $\rm \feh$ value for the star. The index and its calibration using known globular cluster giants is described extensively in Morrison et al. (in prep.). It utilizes the strong luminosity sensitivity of the Mg Ib triplet and MgH features near 5200 $\rm \AA$\footnote{There is a similar index in the SSPP output, but this is based on continuum-corrected data. The continuum correction actually removes some of the signal from the strong MgH bands in K dwarfs, so this is not as sensitive as our index.}.

We need to keep track of the different targeting categories for our sample stars, as their metallicity distributions differ significantly. A complication is introduced by the fact that the SDSS photometry has been continually improved between the start of the SEGUE project in 2005 and Data Release $\rm 8$ in 2011. Because of the slight changes in $g-r$ and $u-g$ colors during this time, stars targeted using earlier photometry may not satisfy the criteria for target selection if one uses the most recent photometry. Note that we do not use kinematic selection criteria in our target selection, except for the proper-motion cut described above, which only affects giants with high velocity at very close distances (see Morrison et al. in prep., for more discussion.). This group includes stars targeted originally as K giants whose new photometry moved them out of the target boxes, and also stars targeted originally in other categories.

Using colors, reddening, log g values, and spectra
  from DR9, we find 15,750 field-star candidates that satisfy our K-giant
  criteria, have good photometry (i.e., color errors from SDSS pipeline are less than
  0.04 mag), and have passed the Mg Ib triplet and MgH features criterion. We
  describe a further culling of the sample in \S 3.4, aimed at eliminating stars
  that could either be on the RGB or in the red clump. The error of $\rm \feh$ for each K giant
  used in this paper is calibrated using cluster data plus repeat
  observations, which depends on the signal-to-noise ratio of the
  spectrum\footnote{The relation between the error of $\feh$ ($\Delta\feh$) and the signal-to-noise ($\rm SN$) ratio is expressed as $\rm \Delta\feh=\sqrt{0.07^2+(0.48-0.02SN+4\times10^{-4}SN^2-2.4\times10^{-6}SN^3)^2}$ for $\rm 17<SN<66$; the out-of-range SN values are truncated to the nearest value of $\rm \Delta \feh$.}, as described in detail in Morrison et al. (in prep.).

We show below that it is important to well-quantify the errors in color measurement. The two contributing factors here are the measurement errors on the $g-r$ color and on the reddening E(B-V). While the SDSS PHOTO pipeline gives estimates of measurement error on each color, these estimates do not include effects such as changes in sky transparency, mis-matches between the model used for the point spread function and the actual stellar image, and so on. 

We estimate this additional factor as $\rm 0.011$ magnitudes in $g-r$ \citep{Padmanabhan2008}. To quantify the reddening errors, one of the authors (HLM) has selected 102 globular clusters from the compilation of \citet[][2010 edition]{Harris1996} with good color magnitude diagrams in the literature, and compared the estimates of E(B-V) from \citet[][SFD hereafter]{Schlegel1998} with those of Harris. Here we assume that the globular cluster reddenings represent ``ground truth" for Galactic structure studies. We find that for objects with E(B-V) from SFD less than 0.25 mag (our limit for the K-giant investigation) there is a small offset (SFD reddenings are on average 0.01 mag higher than those of Harris). Assuming that both error estimates contribute equally to the differences between them, we find an error for the SFD reddenings of 0.013 mag. 

Thus, to account for both of these effects, we add 0.017 magnitudes in quadrature to the estimate of $g-r$ error, and 0.037 magnitudes in quadrature to the estimate of the $r$ error from the SDSS pipeline.

\section{Probabilistic Framework for Distance Estimates}
Our goal is to obtain the {\it posterior probability distribution
  function} (pdf) for the distance modulus of any particular K-giant
star, after accounting for (i.e., marginalizing over) the
observational uncertainties in apparent magnitudes, colors, and
metallicities ($\rm {\it m}, ~{\it c}, ~\feh,~\Delta {\it m},\\~\Delta
{\it c},~\Delta \feh$), and after including available prior
information about the K-giant luminosity function, metallicity
distribution, and, possibly, the halo radial density profile.

\subsection{Distance Modulus Likelihoods}

We start by recalling Bayes theorem, cast in terms of the situation at
hand:
\begin{equation}
\rm P\bigl(\DM\mid \{\m,\co,\feh\}\bigr)=\frac{P\bigl(\{\m,\co,\feh\}\mid \DM
  \bigr)p_{prior}\bigl(\DM \bigr)}{P\bigl(\{\m,\co,\feh\}\bigr)}.
\end{equation}\\
Here, $\rm P\bigl(\DM \mid \{\m,\co,\feh\}\bigr)$ is the pdf of the distance moduli, ($\rm \DM=m-M$), and describes the
relative probability of different $\rm \DM$, in light of the data,
\{$\m,\co,\feh$\} (we use \{ \} to denote the observational
constraints, i.e., the estimates and the uncertainties for the
observable quantities). $\rm P\bigl(\{\m,\co,\feh\} \mid \DM \bigr)$ is
the {\it likelihood} of $\rm \DM$ (e.g., $\rm \ldm$), and tells us how
probable the data \{$\m,\co,\feh$\} are if $\rm \DM$ were true. The term $\rm
p_{prior}\bigl(\DM \bigr)$ is the {\it prior probability} for the $\DM$,
which reflects independent information about this quantity, e.g., that
the stellar number density in the Galactic halo follows a power law of
$\rm r^{-3}$. The term $\rm P\bigl(\{\m,\co,\feh\}\bigr)$ is a non-zero
constant.

So, the probability of the $\rm \DM$ for a given star is proportional
to the product of the likelihood of $\rm \DM$ and the {\it prior
  probability} of $\rm \DM$ (e.g., Eq 2).

\begin{equation}
\rm P\bigl(\DM \mid \{\m,\co,\feh\}\bigr)\propto \ldm p_{prior}\bigl(\DM
\bigr)
\end{equation}

The {\it prior probability} for the $\rm \DM$ in Eq. 2 can be incorporated
independently, and the main task is to derive $\rm \ldm$. In deriving
$\rm \ldm$, we must in turn incorporate any {\it prior} information
about other parameters that play a role, such as the giant-branch
luminosity function, $\rm p_{prior}\bigl({\rm M}\bigr)$, or the
metallicity distribution of halo giants, $\rm
p_{prior}\bigl(\feh\bigr)$. This is done via:

\begin{equation}
\rm \ldm =\int\int p\bigl(\{\m,\co,\feh\}\mid \DM,{\rm
  M,FeH}\bigr)p_{prior}\bigl({\rm M}\bigr)p_{prior}\bigl({\rm
  FeH}\bigr)d{\rm M}~d{\rm FeH}.
\end{equation}

\noindent Here we use FeH to denote the metallicity of the model, while we use
$\rm \feh$ for the observed metallicity of the star.

\subsection{Observables and Priors}
The direct observables we obtain from SEGUE and the SSPP are the
extinction corrected apparent magnitudes, colors, metallicities, and
their corresponding errors $\rm \bigl({\it r_0},~{\it
  (g-r)_0},\\~\feh,~\Delta {\it r_0},~\Delta {\it (g-r)_0},~\Delta
\feh)$. Figure~\ref{f:fcolor} shows the color-color diagram for our confirmed K giants, following application of the procedures described below. Figure~\ref{f:fsample} shows that the most common stars are
the intrinsically fainter blue giants, as we would expect from the
giant-branch luminosity function. Hereafter, we use $c$ and $m$
instead of $(g-r)_0$ and $r_0$ for convenience and generality in the
expression of the formulas.

In our analysis, we can and should account for three pieces of prior
(external) information or knowledge about the RGB population that go
beyond the immediate measurement of the one object at hand: $\rm
p_{prior}(\DM),~p_{prior}\bigl(M\bigr)$, and $\rm
p_{prior}\bigl(\feh\bigr)$ (shown as Eq. 1 and Eq. 3).

The {\it prior probability} of $\rm \DM$ reflects any information on
the radial density profile of the Milky Way's stellar
halo. \citet{Vivas2006} and \citet{Bell2008} indicated that the radial
halo stellar density follows a power law $\rm \rho (r) \propto
r^\alpha$, with the best value of $\rm \alpha=-3$, and reasonable
values in the range $\rm -2>\alpha>-4$; this implies $\rm
p_{prior}(\DM)d\DM=\rho(r)4\pi r^2dr$, $\rm p_{prior}(\DM) \propto
e^{\frac{(3+\alpha)log_e10}{5}\DM}$. Quite fortuitously, the {\it
  prior probability} for $\rm \DM$ turns out to be flat for the radial
stellar density of a power law of $\rm \rho (r) \propto r^{-3}$. Given
that the $\rm \ldm$ approximatively follows a Gaussian distribution
with the mean of $\rm \DM_0$ and standard deviation of $\rm \Delta
\DM$ (here $\rm \Delta \DM$ is the error of $\rm \DM$), the $\rm
p_{prior}(\DM)$ will shift the estimate of $\rm \DM_0$ by $\rm
\frac{(3+\alpha)log_e10}{5}(\Delta \DM)^2$, but with basically no
change in $\rm \Delta\DM$. Therefore, the shifts in the mean $\rm \DM$
caused by $\rm p_{prior}(\DM)$ can be neglected for values of $\rm
-2>\alpha>-4$ (i.e. $\rm \frac{(3+\alpha)log_e10}{5}(\Delta\DM)^2 \ll
\rm \Delta \DM$). In \S 3.6, we explicitly verify that different
  halo density profiles do not affect the distance estimation
  significantly using artificial data.

The {\it prior probability} of the absolute magnitude M can be
inferred from the nearly universal luminosity function of the giant
branch of old stellar populations. Specifically, we derive it from the
globular clusters $\rm M5$ ($\rm \feh=-1.4$) and $\rm M30$ ($\rm
\feh=-2.13$) \citep{sandquist1996,sandquist1999}, and from the Basti
theoretical luminosity function with $\rm \feh=-2.4$ and $\rm \feh=0$
\citep{Pietrinferni2004}. Figure~\ref{f:flprior} (top panel) shows
that the luminosity functions for the RGBs derived from the globular
clusters in the different bands are consistent with one another, and
also with the Basti theoretic luminosity functions for the
metal-rich and metal-poor cases. All the luminosity functions follow
linear functions with similar slope, $\rm k=0.32$, as a result of the fact that
the luminosity functions are insensitive to changes in the metallicity and color. According to $\rm p(M)dM=p(L)dL$ and $\rm M \sim -2.5
log L$, the luminosity function $\rm p(M) \propto 10^{kM}$ means $\rm
p(L) \propto L^{-2.5k-1}$. We conclude that the luminosity function
for the giant branch follows $\rm p(L) \propto L^{-1.8}$, shown as the
dashed line in Figure~\ref{f:flprior}.

Our {\it prior probability} for $\rm \feh$ results from an empirical
approach. In the SEGUE target selection, the K giants were split into
four sub-categories: the l-color K giants, the red K giants, the proper-motion K
giants, and the serendipitous K giants. This suggests that one should
adopt the overall metallicity distribution of each sub-category as the
$\rm \feh$ prior for any one star in this sub-category
(Figure~\ref{f:ffehprior}). Figure~\ref{f:ffehprior} shows the
metallicity distribution variation with apparent magnitude in the upper
panel, and the four $\rm \feh$ priors in the lower panel. It can be
seen that the four sub-categories have different metallicity
distributions. For a star that has approximately the mean metallicity,
this prior should leave $\rm \ldm$ unchanged, because the individual
metallicity error is smaller than the spread of the $\rm
p_{prior}\bigl(\feh\bigr)$. However, for a star of seemingly very low
metallicity, the prior implies that this has been more likely an underestimated metallicity of a (intrinsically) less metal-poor star, which would lead to an overestimated distance modulus.

\subsection{Color-Magnitude Fiducials}

To obtain distance estimates, we determine an estimate of the absolute
magnitude of each star, using its $(g-r)_0$ color and a set of
giant-branch fiducials for clusters with metallicities ranging from
$\rm \feh=-2.38$ to $\rm \feh=+0.39$, and then use the star's apparent magnitude
\citep[corrected for extinction using the estimates of][]{Schlegel1998}
to obtain its distance. We prefer to use {\sl fiducials}, rather than
{\sl model-isochrone} giant branches wherever possible, because
isochrone giant branches cannot reproduce cluster fiducials with
sufficient accuracy.

As the SDSS imager saturates for stars brighter than $g \sim 14.5$, almost none of the clusters observed by SDSS and used by \citet{An2008} have unsaturated giant branches. Therefore, we
  derived such fiducials, using the globular clusters M92, M13, and
  M71, and the open cluster NGC6791, which have accurate
  $ugriz$ photometry from the DAOphot reductions of \citet{An2008} for most of the stars; the giant branches can also be supplemented using the $u'g'r'i'z'$ photometry of \citet{Clem2008}. We transformed to
$ugriz$ using the transformations given in \citet{Tucker2006}. Note that M71 is
  a disk globular cluster, and one of the few clusters in the North at
  this important intermediate metallicity. However, it has reddening that varies somewhat over the face of the cluster, making it a difficult cluster to work with. We use $g-i$ instead of $g-r$ to obtain more accurate estimates of the variable reddening map of M71, and use them to produce a better fiducial for this cluster (see Morrison et al. in prep. for details). We list our adopted values of $\rm \feh$, reddening, and distance modulus for each cluster in Table 1. In addition, we supplemented the fiducials with a Solar-metallicity giant branch from the Basti $\alpha$-enhanced isochrones \citep{Pietrinferni2004}. Figure~\ref{f:ffiducial} shows the four fiducials and the one theoretical isochrone. The color at a given $\rm M$ and $\rm \feh$, $\rm {\it c}\bigl(M,\feh\bigr)$, can then be interpolated from these color-magnitude fiducials.

It is important to note that most of the halo K giants are $\alpha$-enhanced, except for a few giants close to Solar metallicity (Morrison et al. in prep.). For less metal-poor K giants, the effect of $[\alpha/Fe]$ on luminosity is stronger ({\it for instance, the difference of r-band absolute magnitude can be as large as 0.5 mag at the tip of the giant branch for an $\alpha$-enhanced giant compared to one with Solar-scaled $\alpha$-element abundance but the same $\rm \feh$ value}). The cluster fiducials and one isochrone we use for distance estimates when $\rm \feh<0$ are $\alpha$-enhanced, while above that metallicity, we assume gradual weakening of the $\alpha$-abundance, naturally introduced by the NGC6791 (Solar-scaled alpha abundances) fiducial line in the interpolation. In other word, when a giant's $\rm \feh$ is between Solar and the NGC6791 value, its $\alpha$-abundance is also assumed to be in between.

Given the sparse sampling of the M -- $(g-r)_0$ space by the four
isochrones, we need to construct interpolated fiducials. We do this by
quadratic interpolation of $\rm {\it c}\bigl(M,\feh\bigr)$, based on
the three nearest fiducial points in color, and construct a dense
color table for given $\rm M$ and $\rm \feh$, which will be used for
Eq. 4. Extrapolation beyond the metal-poor and metal-rich boundaries
and the tip of the RGB would be poorly constrained. Therefore, we use
these limiting fiducials instead for the rare cases of stars with $\rm
\feh<-2.38$ or $\rm \feh>+0.39$. Table 3 shows an interpolated
fiducial with $\rm \feh=-1.18$; the entire catalog of 20 interpolated
fiducials with metallicity ranging from $\rm \feh=-2.38$ to $\rm \feh=+0.39$ is
provided in the electronic edition of the journal. While there is more
than one way to interpolate the colors, such as quadratic or piecewise
linear, we have checked and found that different interpolation schemes
lead to an uncertainty less than $\rm \sim \pm 0.02$ mag in color, due
to the sparsity of the fiducials, which could be an additional source of error in $\DM$ estimates.

\subsection{Red Giant-Branch Stars vs. Red-Clump Giants }
\begin{figure}
\includegraphics[width=\textwidth]{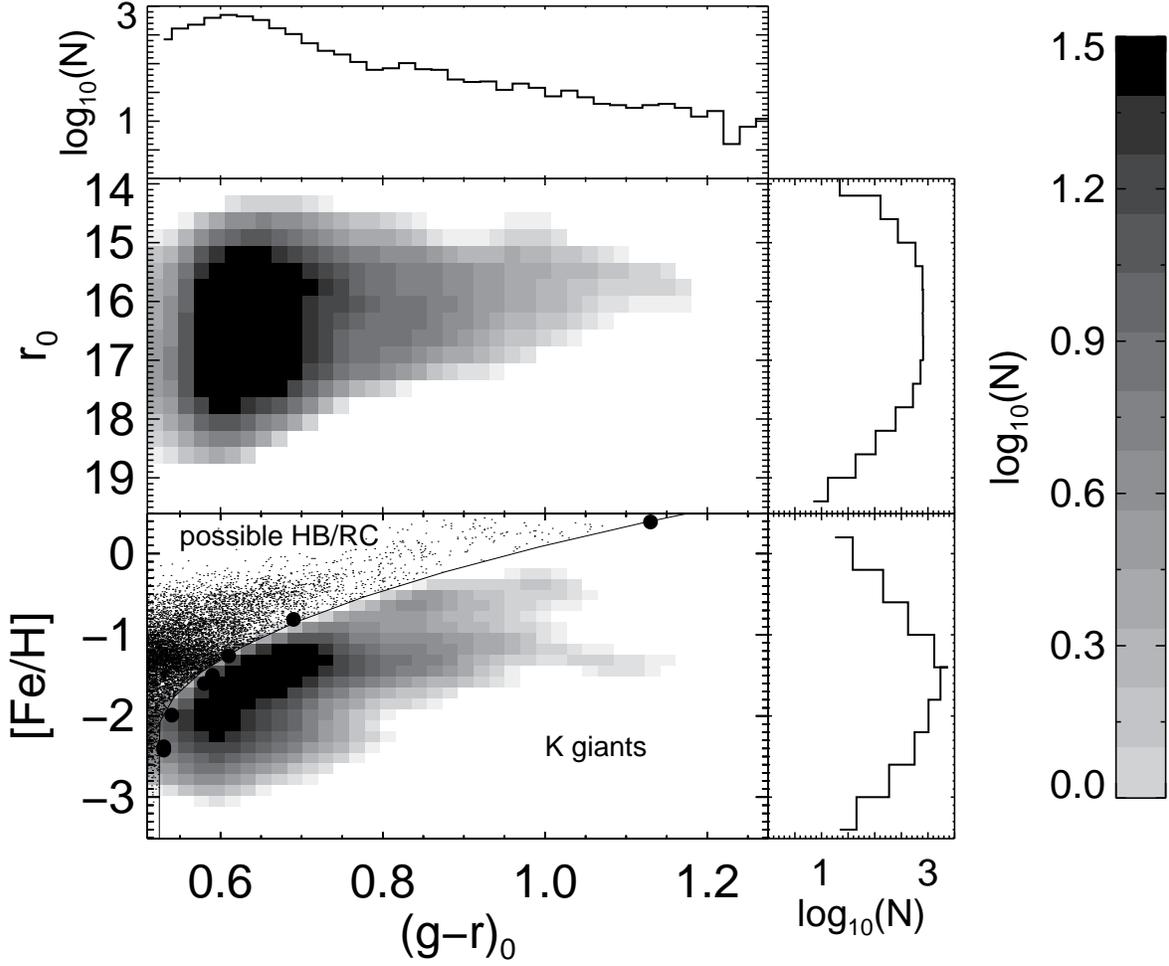}
\caption{Distribution of our K-giant sample in the color-magnitude and
  color-metallicity plane. It can be seen that the most common stars
  are the intrinsically fainter blue giants, as we would expect from
  the giant-branch luminosity function. The possible HB/RC stars are
  over-plotted as dots. The filled circles are the observed points
  drawn from clusters published by \citet{An2008}, from which the
  relation between $(g-r)_0^{HB}$ and $\rm \feh$ (solid line) was
  derived.}
\label{f:fsample}
\end{figure}

\begin{figure}
\includegraphics[width=\textwidth]{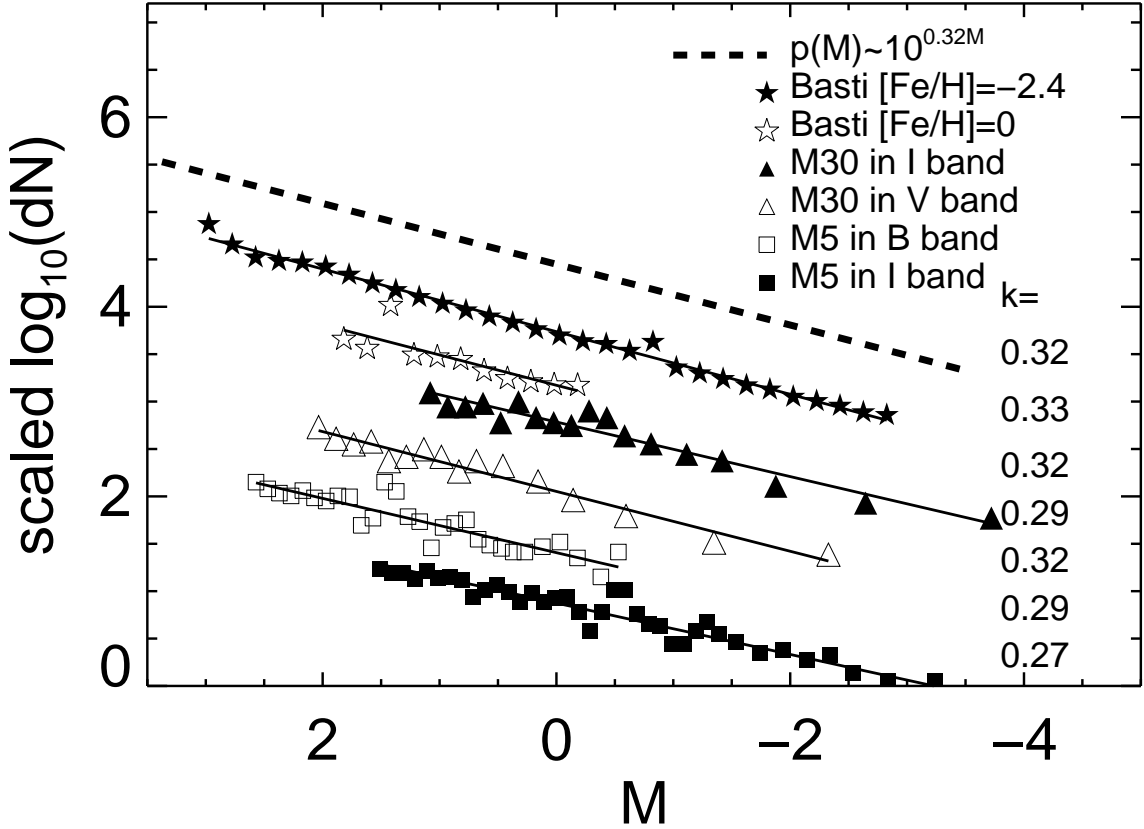}
\caption{Luminosity functions for the giant branches of two globular
  clusters in different bands, compared with two theoretical
  giant-branch luminosity functions. We scaled $\rm log_{10}(dN)$ to
  make the luminosity functions separate from each other, because the
  slope is the important parameter to test whether the luminosity
  functions are consistent. All the luminosity functions follow power
  laws with nearly the same slope of k=0.32, so that the theoretical and
  observational luminosity functions are consistent, and both are
  insensitive to changes in metallicity and passbands. Therefore, the
  prior probability adopted for the absolute magnitude in the analysis
  is $\rm p(M)=10^{0.32M}/17.788$, which is based on a variety of
  theoretical and empirical giant-branch luminosity functions, and
  whose integral over [--3.5,+3.5] has been normalized to unity.}
\label{f:flprior}
\end{figure}

\begin{figure}
\centering
\includegraphics[width=0.85\textwidth,height=0.4\textheight]{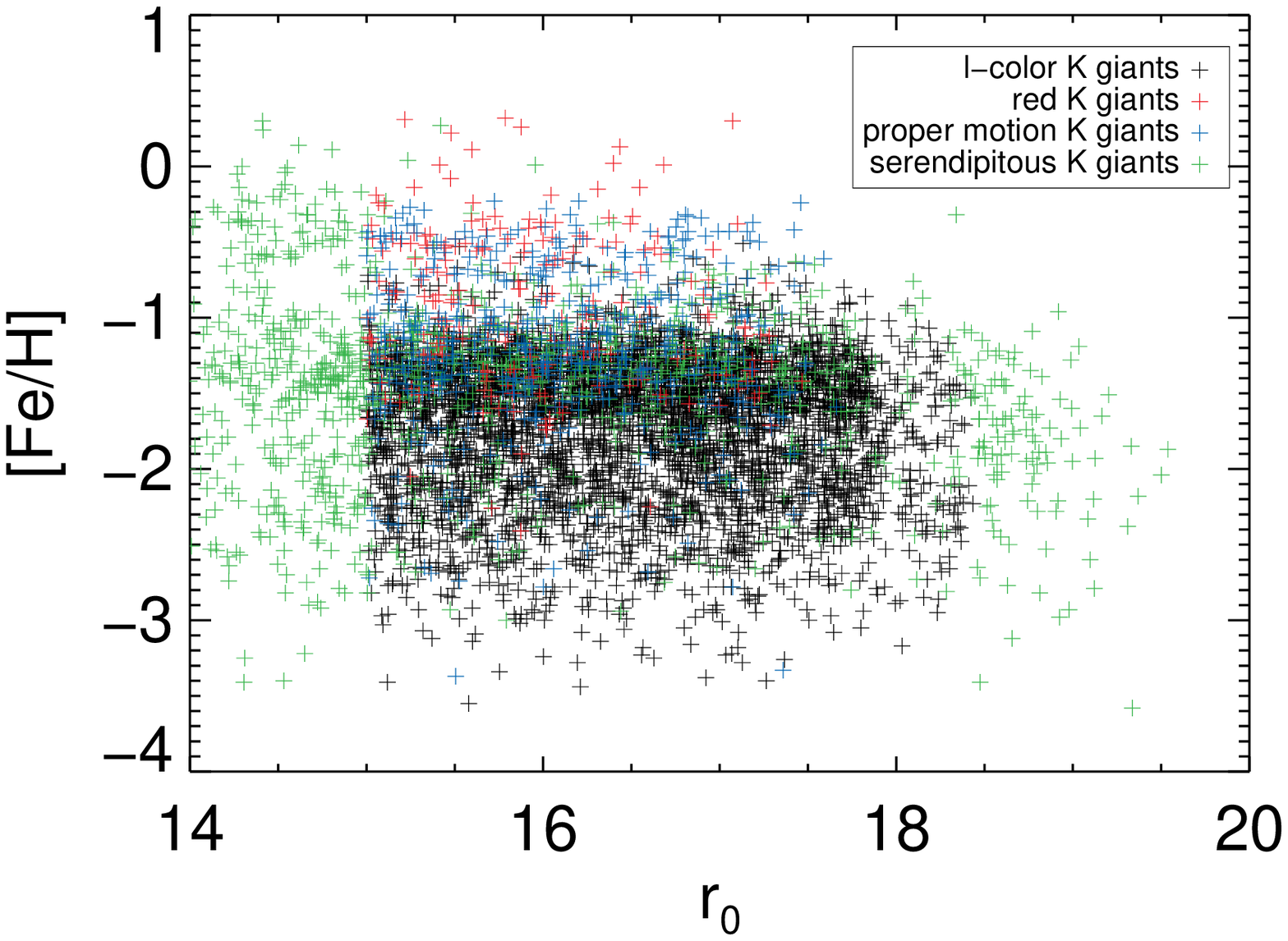}
\includegraphics[width=\textwidth,height=0.6\textheight]{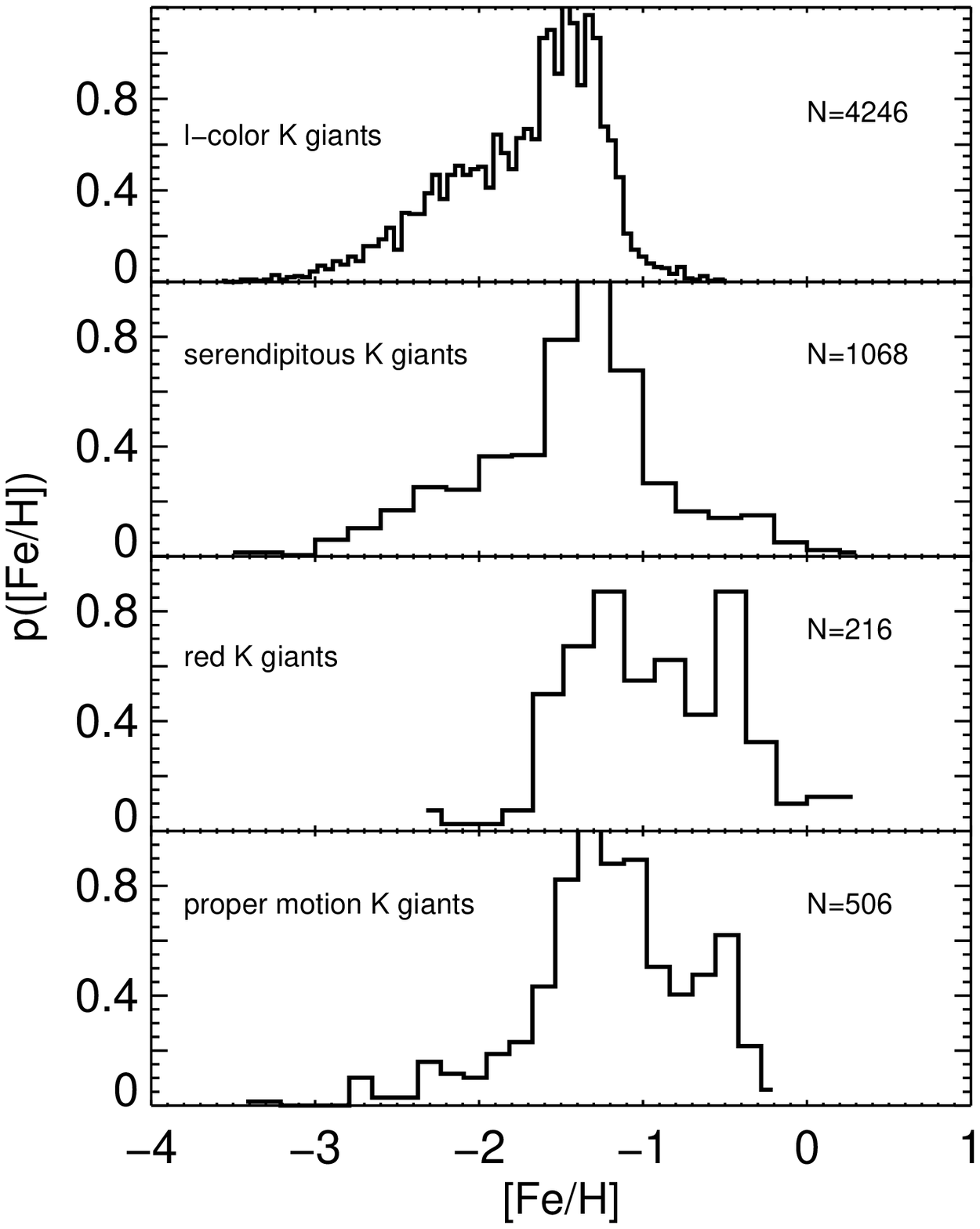}
\caption{(Upper panel) Variation of the metallicity distribution with
  apparent $r_0$ magnitude for four sub-categories. (Lower panel) The
  four $\rm \feh$ priors adopted in this work. The integral of $\rm
  p(\feh)$ has been normalized to unity.}
\label{f:ffehprior}
\end{figure}

\begin{figure}
\includegraphics[width=\textwidth]{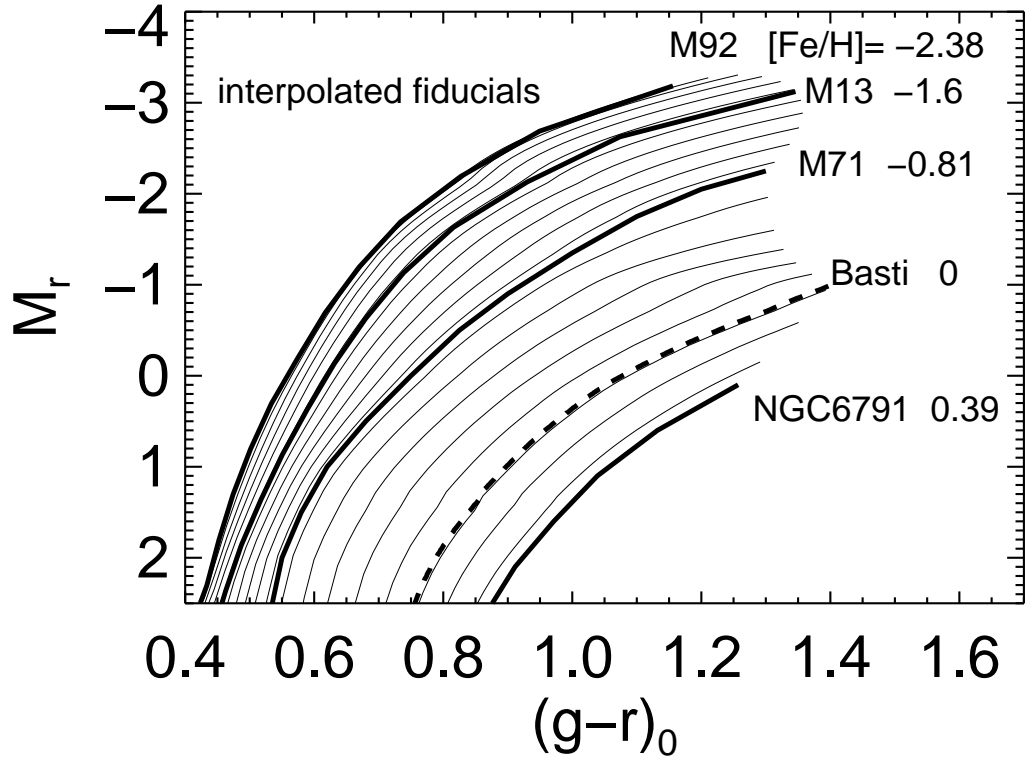}
\caption{Interpolation of the four giant-branch fiducials (thick
  lines), to obtain the relation, $(g-r)_0=f(M_{\it r},\feh)$, for any
  set of $\rm (M_{\it r},\feh)$. The thin lines show a set of
  interpolated fiducials. No values outside the extreme fiducials are
  used.}
\label{f:ffiducial}
\end{figure}

In addition, we have chosen not to assign distances to stars that lie
on the giant branch below the level of the horizontal branch
(HB). This is because the red horizontal-branch or red-clump (RC)
giants in a cluster have the same color as these stars, but quite
different absolute magnitudes, and the SSPP log g estimate is not
sufficiently accurate to discriminate between the two options. We
derive a relation between $\rm \feh$ and the $(g-r)_0$ color of the
giant branch at the level of the horizontal branch, using eight
clusters with $ugriz$ photometry from \citet{An2008}, with cluster
data given in Table 2. The $\rm \feh$ and $(g-r)_0^{HB}$ for the HB/RC
of the clusters follow a quadratic polynomial,
$(g-r)^{HB}_0=0.087\feh^2+0.39\feh+0.96$, as shown in
Figure~\ref{f:fsample}. We then use this polynomial and its $\rm \feh$
estimate to work out, for each star, whether it is on the giant branch
above the level of the HB. It turns out that more than half of the
ucandidate K giants fall into the region of RGB - HB ambiguity. 

To incorporate the errors of metallicities and colors, we envisage each star as a 2D (error-) Gaussian in the color-metallicity plane, centered on its most likely value and the 2D Gaussian having widths of color errors and metallicity errors, respectively. Then, we can calculate the ``chance of being clearly RGB'' as the fraction of the 2D integral over the 2D error-Gaussian that is to the right of the line. 

Ultimately, we are left with 6036 stars with more than 45\% chance of being clearly on
the RGB, above the level of the HB. Of these, 5962 stars have more than 50\% chance to be RGB, 5030 stars have more than 68\% chance to be RGB, and 3638 stars have more than 90\% chance to be RGB. In addition, 216 satisfy the target criteria for red
K giants, 506 the criteria for proper-motion K giants, and 4246 the
l-color K-giant criteria. Another 1068 were serendipitous
identifications -- stars targeted in other categories which nevertheless
were giants. Figure~\ref{f:fsample} shows the distribution of the
apparent magnitudes, $\rm r_{0}$, and metallicities, $\rm \feh$, along
with the color, $(g-r)_0$.

Besides the contamination from HB/RC stars, we need to consider
possible contamination of our sample by asymptotic giant-branch (AGB)
stars, because it is not possible for us to distinguish RGB
from AGB stars with our spectra. While the difference in absolute
magnitude can be large (reaching $\sim$ 0.8 mag, implying a 40\%
distance underestimate at the blue end of our giant color range), the proportion of our
giants that are on the AGB is relatively small. We used both the
luminosity function of \citet{sandquist2004} for the globular cluster
M5 and evolutionary tracks from Basti isochrones for old populations
of metallicity $\rm \feh=-2.6$ and $\rm \feh=-1.0$, to estimate the percentage of stars
which are on the AGB. We find that for the most metal-poor stars,
around 10\% will be AGB stars, while for stars with $\rm \feh$ close
to $\rm \feh=-1.0$ the fraction is near 20\%. For less metal-poor stars
  with $\rm \feh >-1.0$, the expected fraction of AGB stars becomes larger than 20\%,
  but in the SEGUE K-giant sample, less than 10\% of stars have $\rm \feh>-1.0$.

\subsection{Implementation}

For any given star, the observables are its apparent magnitude and
associated error, $\rm \bigl(\m_i,\Delta \m_i\bigr)$, its color and
error $\rm \bigl(\co_i,\Delta \co_i\bigr)$, and its metallicity and
error $\rm \bigl(\feh_i,\Delta \feh_i\bigr)$. The $\rm \DM$ and the
data are linked through the absolute magnitude M via: $\rm
\m_i=M+\DM_i$ and the fiducial $\rm \co\bigl(M,FeH\bigr)$, which we
presume to be a relation of negligible scatter. Now we can incorporate
the errors of the data and the specific priors on the stellar
luminosity and metallicity distribution when calculating $\rm \ldm$
(see Eq. 3).

In practice, the errors on color, apparent magnitude, and metallicity
can be approximated as Gaussian functions, in which case $\rm
p\bigl(\{\m,\co,\feh\}\mid \DM,M,FeH\bigr)$ (see Eq. 3) is modeled as
a product of Gaussian distributions with mean and Delta $\rm
\bigl(\co_i,\Delta \co_i\bigr)$, $\rm \bigl(\m_i,\Delta \m_i\bigr)$,
and $\rm \bigl(\feh_i,\Delta \feh_i\bigr)$:

\begin{equation}
\begin{split}
\rm p\bigl(\{\m,\co,\feh\}_i\mid
\DM,M,FeH\bigr)=\frac{1}{\sqrt{2\pi}\Delta
  \co_i}\exp\biggl(-\frac{(\co(M,FeH)-\co_i)^2}{2(\Delta
  \co_i)^2}\biggr)\times\\ \frac{1}{\sqrt{2\pi}\Delta
  \m_i}\exp\biggl(-\frac{(\DM+M-\m_i)^2}{2(\Delta
  \m_i)^2}\biggr)\times \frac{1}{\sqrt{2\pi}\Delta
  \feh_i}\exp\biggl(-\frac{(FeH-\feh_i)^2}{2(\Delta\feh_i)^2}\biggr)
\end{split}
\end{equation}
 
For Eq. 3, we use the priors $\rm p_{prior}\bigl(M\bigr)$, based
on the luminosity function of the giant branch, $\rm p\bigl(L\bigr)
\propto L^{-1.8}$ (Figure~\ref{f:flprior}), and $\rm
p_{prior}\bigl(\feh\bigr)$, based on the metallicity distributions of
the K-giant sub-categories (Figure~\ref{f:ffehprior}).

For any K giant with $\rm \{\m_i,\co_i,\feh_i\}$, we can then
calculate $\rm \ldm$ by computing the integral of a bivariate function
(Eq. 3) over dM and dFeH, using iterated Gaussian
quadrature. As described in \S 3.2, the $p_{prior}(\DM)$ is taken as a
  constant for a halo density profile of $\rho(r)\propto
  r^{-3}$, so $\rm P\bigl(\DM \mid \{\m,\co,\feh\}\bigr)\propto
  \ldm$. Then, the best estimate of $\rm \DM$ is at the peak of $\rm
  \ldm$, and its error is the central $\rm 68\%$ interval of $\rm \ldm$
  (i.e. $\rm \frac{\DM_{84\%}-\DM_{16\%}}{2}$).

To speed up the determination of the integral in Eq. 3, we look up
$\rm \co\bigl(M,\feh\bigr)$ in a pre-calculated and finely-sampled color
table, instead of an actual interpolation. This approach can provide a
consistent $\co$ for given $\rm M$ and $\rm \feh$, if the pre-prepared
color table is suitable. We use a color table, $\rm
\co\bigl(M,\feh\bigr)$, of size $\rm 6500 \times 4140$, with $\rm
-3.5<M<3$ and $\rm -3.58<\feh<+0.56$.

\subsection{Tests with Simulated Data Sets}

To test whether our approach leads to largely unbiased $\rm
  \DM$ estimates, a simulated data set was generated in order to mimic the
  SEGUE K-giant sample. As mentioned in \S 2, there are 4
  sub-categories of K giants, and they have different distributions of
  $\rm \feh$, so the simulated stars were generated independently to
  mimic each sub-category. First, we produced a set of
  randomly-generated values of distance, luminosity, and $\rm \feh$,
  following a halo stellar density profile of $\rm \rho(r) \propto
  r^{-3}$, the luminosity function $\rm p(L) \propto L^{-1.8}$, and
  the metallicity distribution of each category of K giants, to cover
  similar ranges of $\rm \bigl(\DM,~M,~\feh\bigr)$ as our sample of K
  giants. Then the apparent magnitudes and colors $\bigl(m,~\co\bigr)$
  were calculated from $\rm \bigl(\DM,~M,~\feh\bigr)$ and
  fiducials. Gaussian errors were added to directly observable
  quantities $\bigl(m,~\co,~\feh\bigr)$, with variances taken from
  observed SEGUE K giants with similar
  $\bigl(m,~\co,~\feh\bigr)$. Finally, a simulated star was accepted
  only if its magnitude and color fall within the selection criteria
  of the pertinent K-giant sub-category.

\begin{figure}
\centering
\includegraphics[width=\textwidth,height=0.5\textheight]{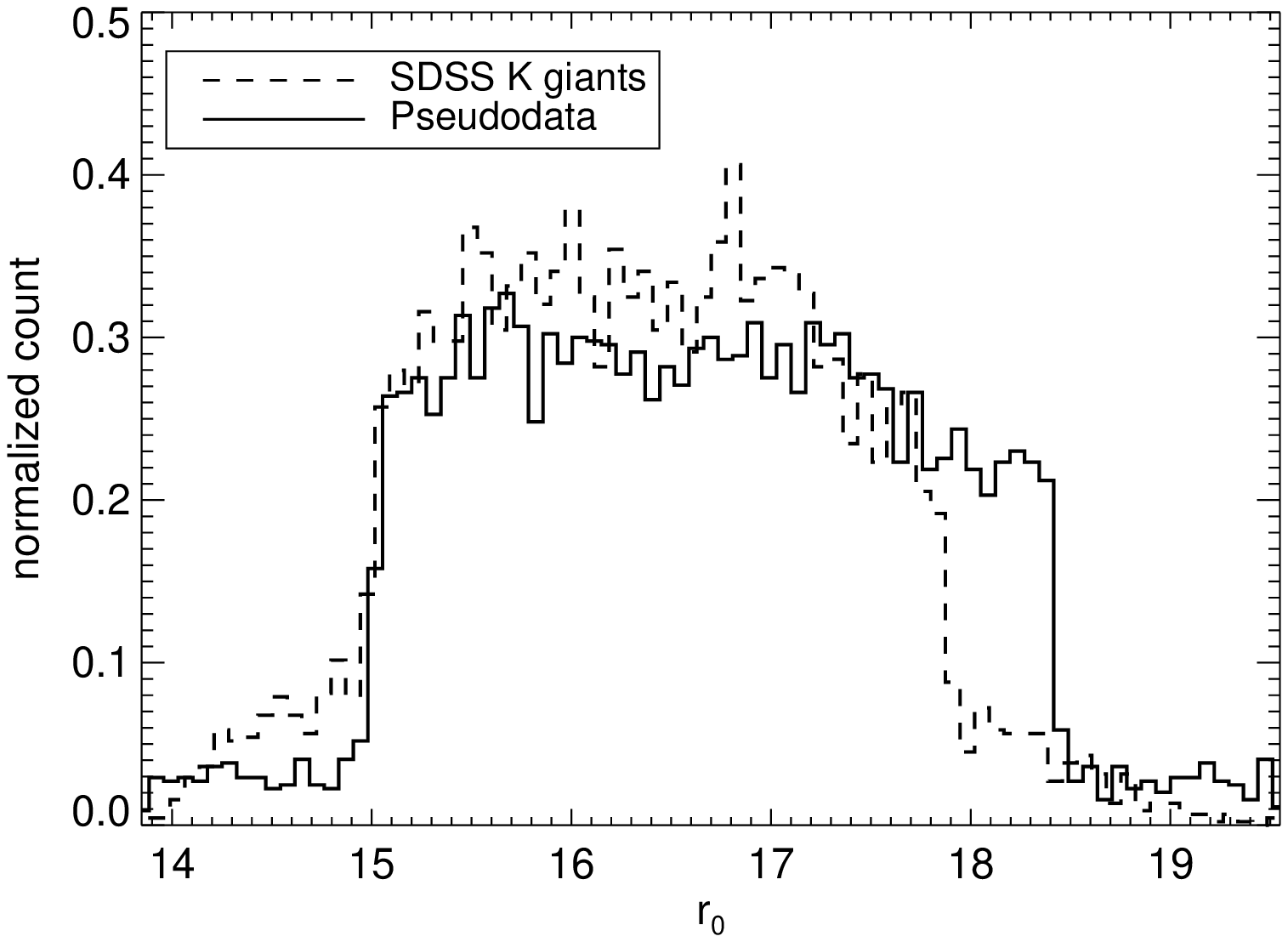}
\includegraphics[width=\textwidth,height=0.5\textheight]{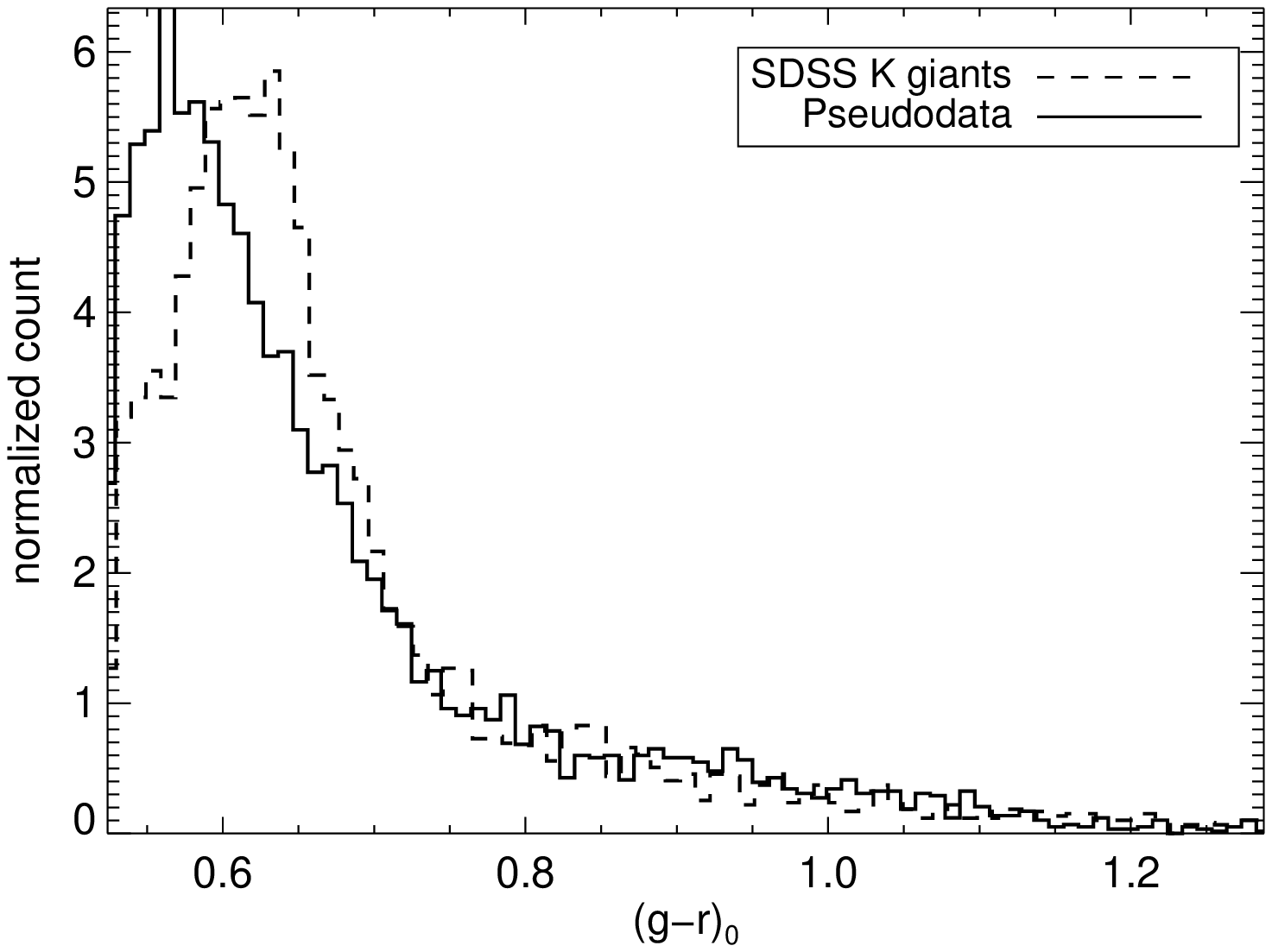}
\caption{Distribution in apparent magnitude $r_0$ (upper panel) and
  color $(g-r)_0$ (lower panel) for the simulated data (full lines)
  and SEGUE K giants (dashed lines). Except for $r_0 \rm >18$ and $(g-r)_0 \rm <0.7$, the distributions are very similar.}
\label{f:test1}
\end{figure}

 A total of 6036 simulated stars were generated according to the above
  procedure. The distributions in $r$ magnitude and in color are
  displayed in Figure~\ref{f:test1}, along with those of SEGUE K
  giants, showing that the simulated sample is a reasonable match,
  except for some apparent incompleteness at the faint end in the
  'real' data\footnote{We quote `real' data to contrast with simulated
    data.}. This sample was then analyzed using the same
  approach to estimate the $\rm \DM$ that was applied to actually observed
  SEGUE K giants, using the known-to-be-correct priors. When
  considering the difference between the calculated distance modulus
  of each star and its true value, divided by the distance modulus
  uncertainty, $\bigl(\rm
  \DM_{cal.}-\DM_{true}\bigr)/\rm \sigma_{\DM_{cal.}}$, we should then
  expect a Gaussian of mean zero and a variance of unity. Indeed, we
  find a mean of $\rm -0.09$ and a variance of $\rm 0.94$ for the case of using luminosity and metallicity priors, but a mean of $\rm -0.14$ and a variance of $\rm 0.95$ for the case of neglecting the metallicity prior, and a mean of $\rm 0.17$ and a variance of $\rm 0.95$ for the case of neglecting the luminosity prior. Note that
  these are in units of $\sigma_{\DM_{cal.}}$, which is typically $\rm
  0.35$ mag; therefore, any systematic biases in distance will be of order
  1\%. Using both priors should lead to unbiased distance estimates.

For the actual SEGUE data the priors are not known perfectly, as we do not know the overall density
  profile of the halo, particularly at large distances
  \citep{Deason2011,Sesar2011}. Previous work indicated that the halo
  radial stellar density follows a power law $\rm \rho (r) \propto
  r^\alpha$, with reasonable values of $\rm -2>\alpha>-4$
  \citep{Vivas2006,Bell2008}. To test the influence of assuming a different $\rm \rho (r)$, we made two sets of simulated data
  following $\rm \rho (r) \propto r^{-2}$ or $\rm \rho (r) \propto
  r^{-4}$, according to the above procedure, and then applied the Bayesian approach to estimate $\rm \DM_{cal.}$ by using a halo stellar density profile of $\rm \rho(r) \propto r^{-3}$, the luminosity function of $\rm p(L)\propto L^{-1.8}$, and the metallicity distribution of each category as priors. Considering the distribution of $\bigl(\rm \DM_{cal.}-\DM_{true}\bigr)/\rm \sigma_{\DM_{cal.}}$, we found a mean of $-0.14$ and a dispersion of $0.95$ for the $\rm \rho (r) \propto r^{-2}$ case and a mean of $-0.08$ and a dispersion of $0.93$ for the $\rm \rho (r) \propto r^{-4}$ case, very similar to the $\rm \rho (r) \propto r^{-3}$
  case, implying that the exact form of the prior for the halo density
  profile does not affect our results significantly. This is
  consistent with \citet{burnett2010}, who also concluded that
  approximate priors in the analysis of a real sample will yield
  reliable results.

\begin{figure}
\includegraphics{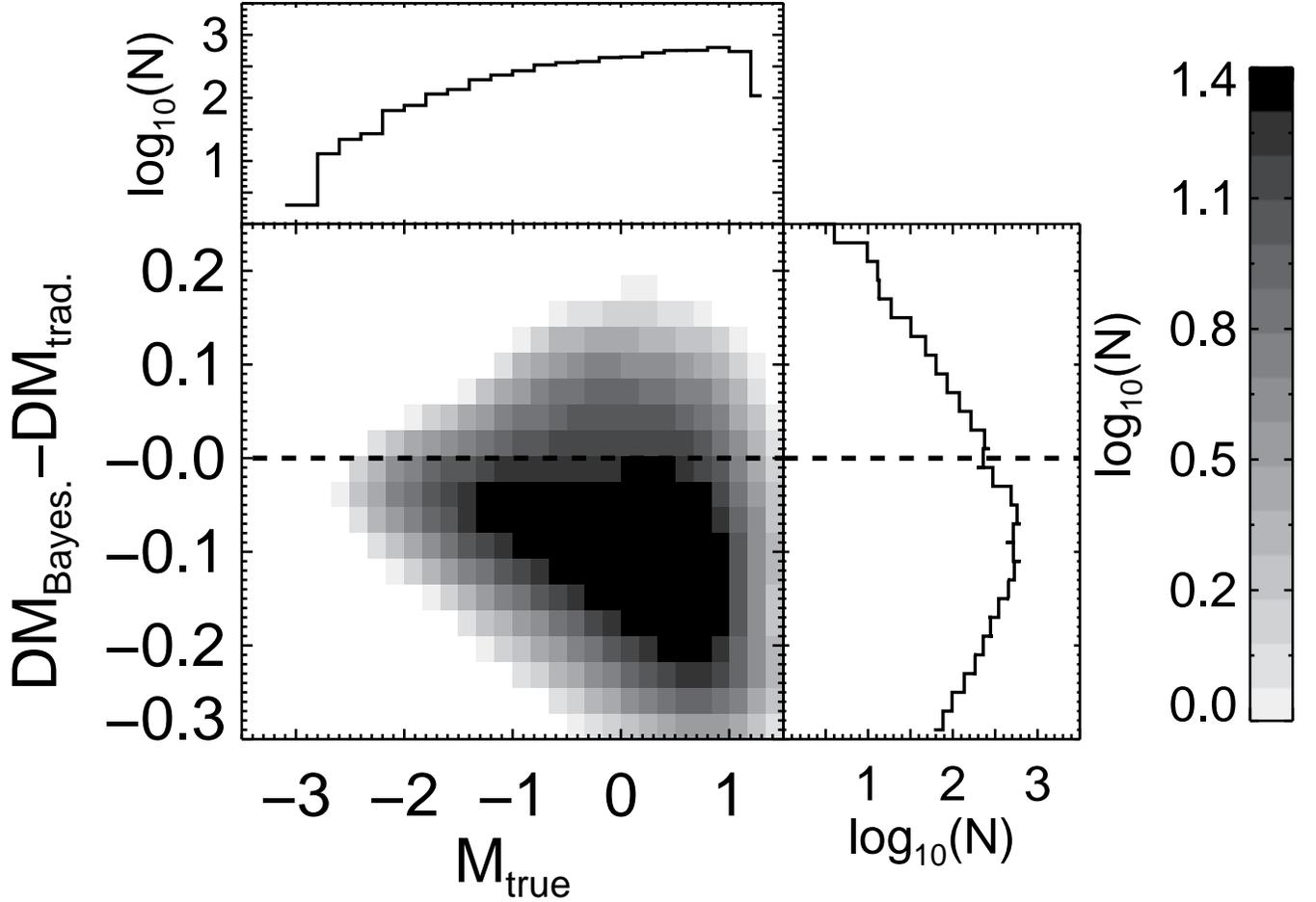}

\caption{The difference between distance moduli estimated by
  traditional and Bayesian methods for the simulated data. Using
  the Bayesian method with luminosity-function prior and metallicity prior
  can help correct a mean overestimate of $\rm \sim 0.1$ mag in the
  distance moduli, compared to the case of neglecting both priors.}
\label{f:fbiassim}
\end{figure}

However, neglecting the luminosity and metallicity priors, as has often been done in previous work
  \citep{Ratnatunga1985,Norris1985,Beers2000}, would lead to a mean systematic distance modulus bias of up 0.1 mag compared to neglecting both priors shown as Figure~\ref{f:fbiassim}, which is also illustrated by the K-giant sample in Figure~\ref{f:fbias} and discussed in \S 4.2. Therefore, only an approach with explicit priors will lead to unbiased distance estimates.

\section{Results}
\subsection{Distances for the SDSS/SEGUE K giants}
\begin{figure}
\includegraphics[width=\textwidth]{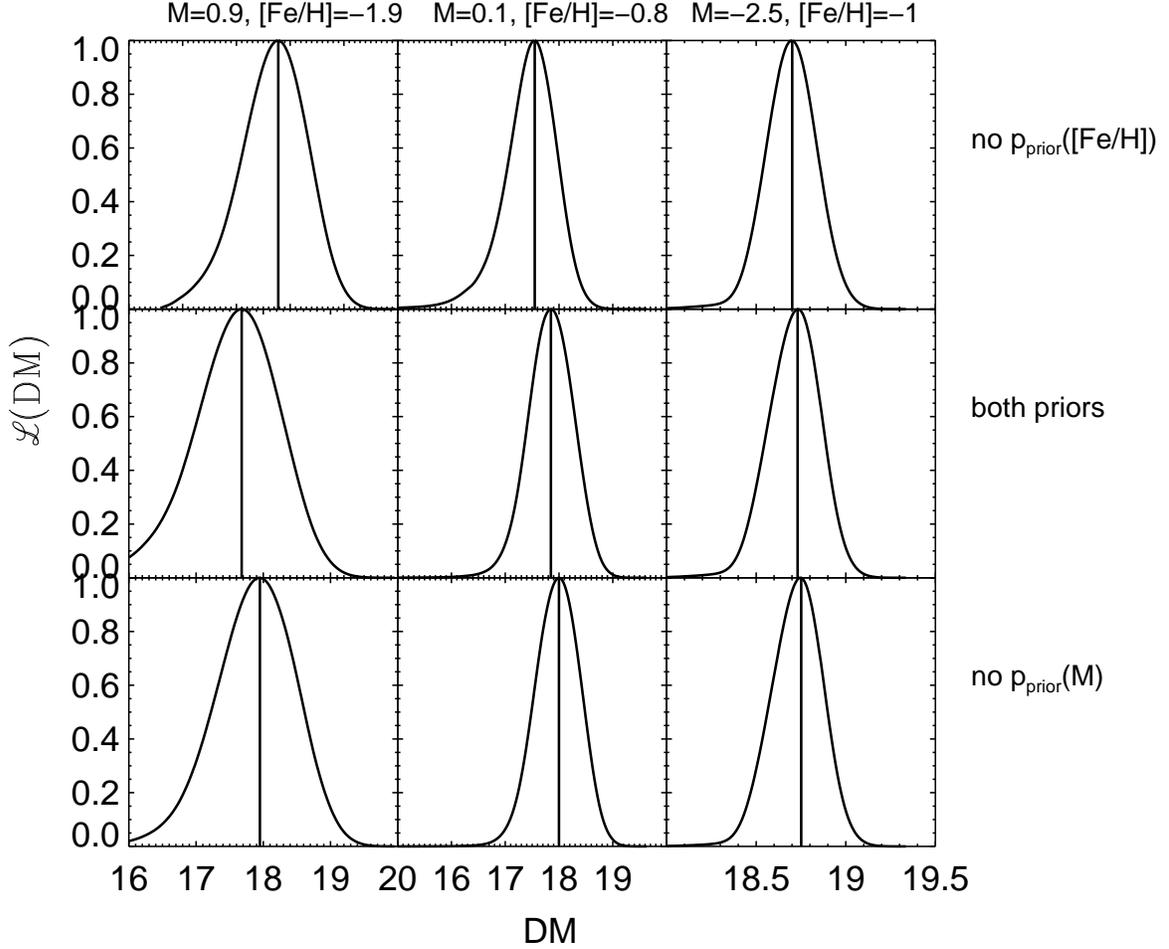}
\caption{Examples of $\rm \ldm$ for three stars, with and without
  accounting for the metallicity and luminosity-function priors (see
  \S 4.2). The black line indicates the most likely $\rm \DM$ under
  the three assumptions. It shows that neglecting luminosity prior leads to the systematic overestimate of $\DM$. As the absolute magnitude increases, the overestimate of the $\DM$ becomes larger. Neglecting the metallicity prior leads to a distance overestimate for metal-poor stars, but a distance underestimate for metal-rich stars.}
\label{f:fldm}
\end{figure}

The most immediate results of the analysis in \S 3.4 are estimates of the distance
  moduli and their uncertainties from $\rm P\bigl(\DM \mid
  \{\m,\co,\feh\}\bigr)$ (Figure~\ref{f:fldm}). At the same time, we obtain estimates for the
  intrinsic luminosities by $\rm M_{\it r}={\it r_0}-\DM_{peak}$, distances from the Sun, and Galactocentric distances by assuming $R_{\sun}=8.0$kpc. This results in the main entries in our public catalog for 6036
  K giants: the best estimates of the distance moduli and their
  uncertainties ($\rm \DM_{peak}$, $\Delta \DM$), heliocentric
  distances, and their errors ($\rm d$, $\Delta d$), Galactocentric
  distances and their errors ($\rm r_{GC}$, $\rm \Delta r_{GC}$), the
  absolute magnitudes along with the errors ($\rm M_{\it r}$, $\rm
  \Delta M_{\it r}$), and other parameters. In addition, we describe
  the $\rm P\bigl(\DM \mid \{\m,\co,\feh\}\bigr)$ by a set of
  percentages, which are also included in the on-line table. Table 4
  shows an example of the online table of K giants. The complete
  version of this table is provided in the electronic edition of the
  journal.

\begin{figure}
\centering
\includegraphics[width=0.7\textwidth,height=0.305\textheight]{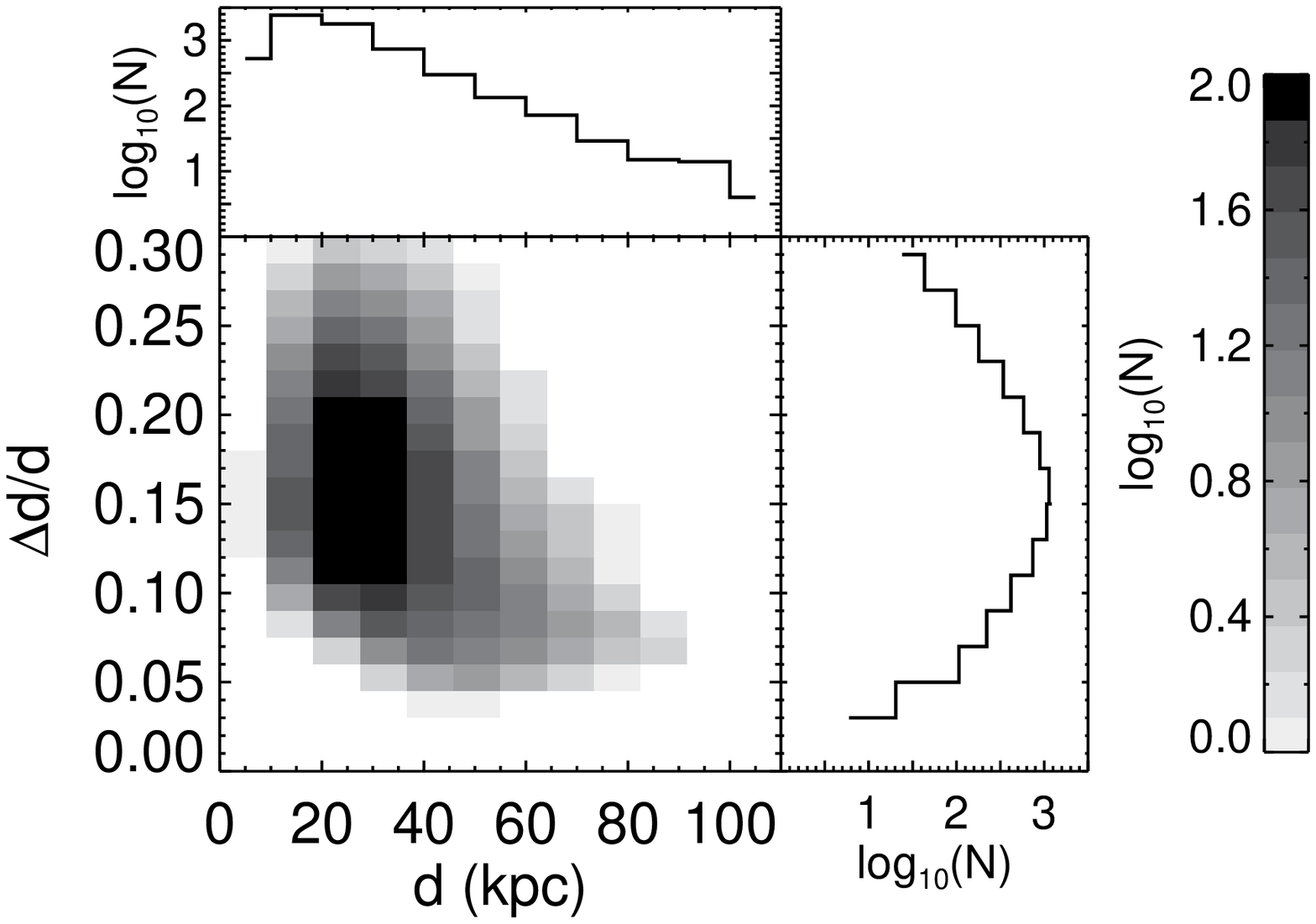}
\includegraphics[width=0.7\textwidth,height=0.305\textheight]{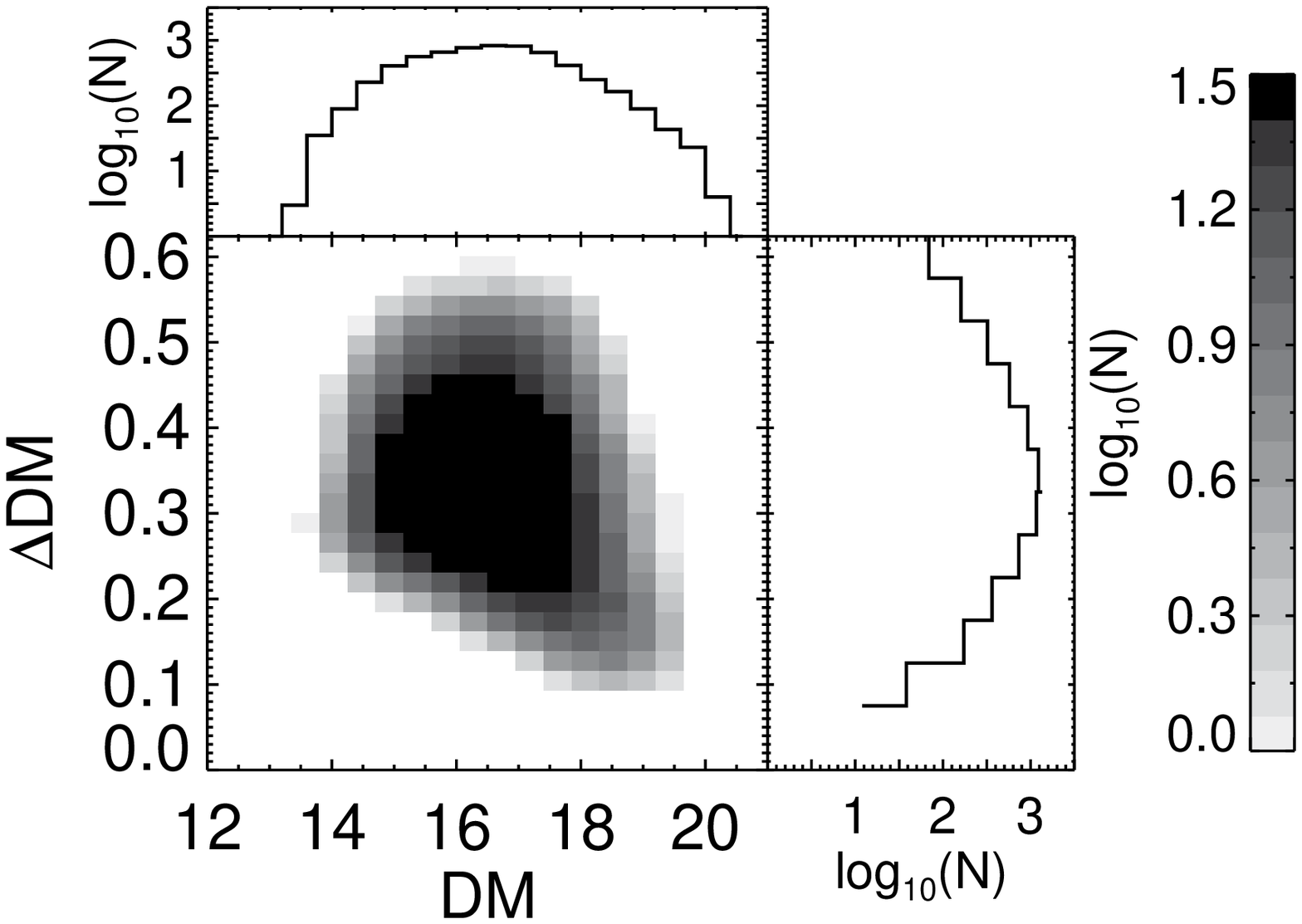}
\includegraphics[width=0.7\textwidth,height=0.305\textheight]{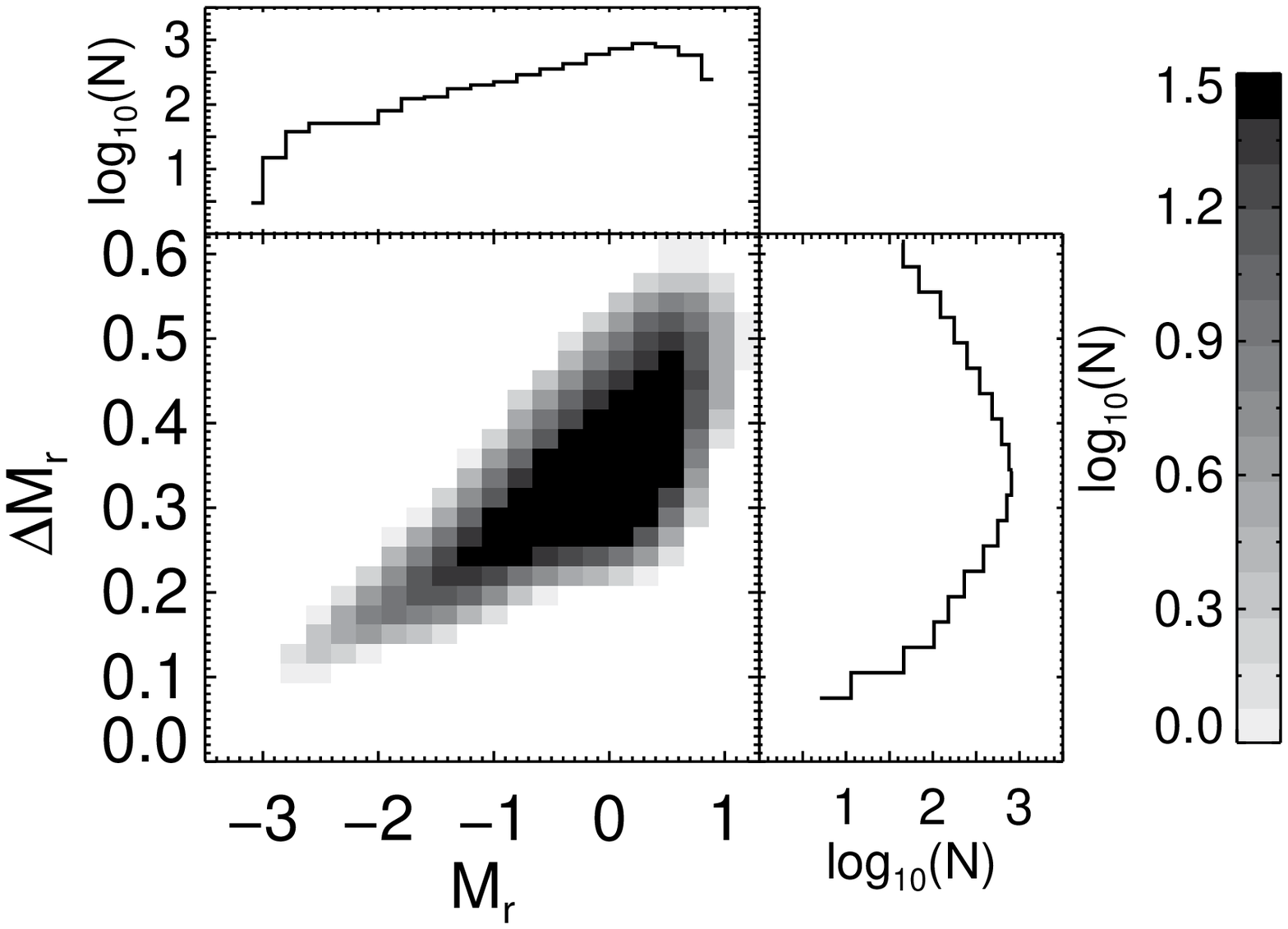}

\caption{Results of the distance estimates for 6036 K giants. (Upper panel) The distribution of the relative errors in distances {\it vs.} distances. (Middle panel) The distribution of the errors in distance moduli {\it vs.} distance moduli. (Lower panel) The distribution of the errors in absolute magnitudes {\it vs.} absolute magnitudes. Note that the fractional distance estimates are less precise for nearby stars, because the lower part of the giant branch (less luminous, therefore more nearby) is steep in the color-magnitude diagram, particularly at low metallicities (see Figure~\ref{f:ffiducial}).}
\label{f:fresult}
\end{figure}

Figure~\ref{f:fresult} illustrates the overall properties of the
ensemble of distance estimates. The top two panels show the mean
precisions of $\rm 16\%$ in $\rm \Delta d/d$ and $\rm \pm 0.35$ mag in
$\rm \DM$; these panels also show that the fractional distances are
less precise for more nearby stars, because they tend to be stars on
the lower part of the giant branch, which is steep in the
color-magnitude diagram, particularly at low metallicities.

\begin{figure}
\includegraphics[width=0.88\textwidth,height=0.45\textheight]{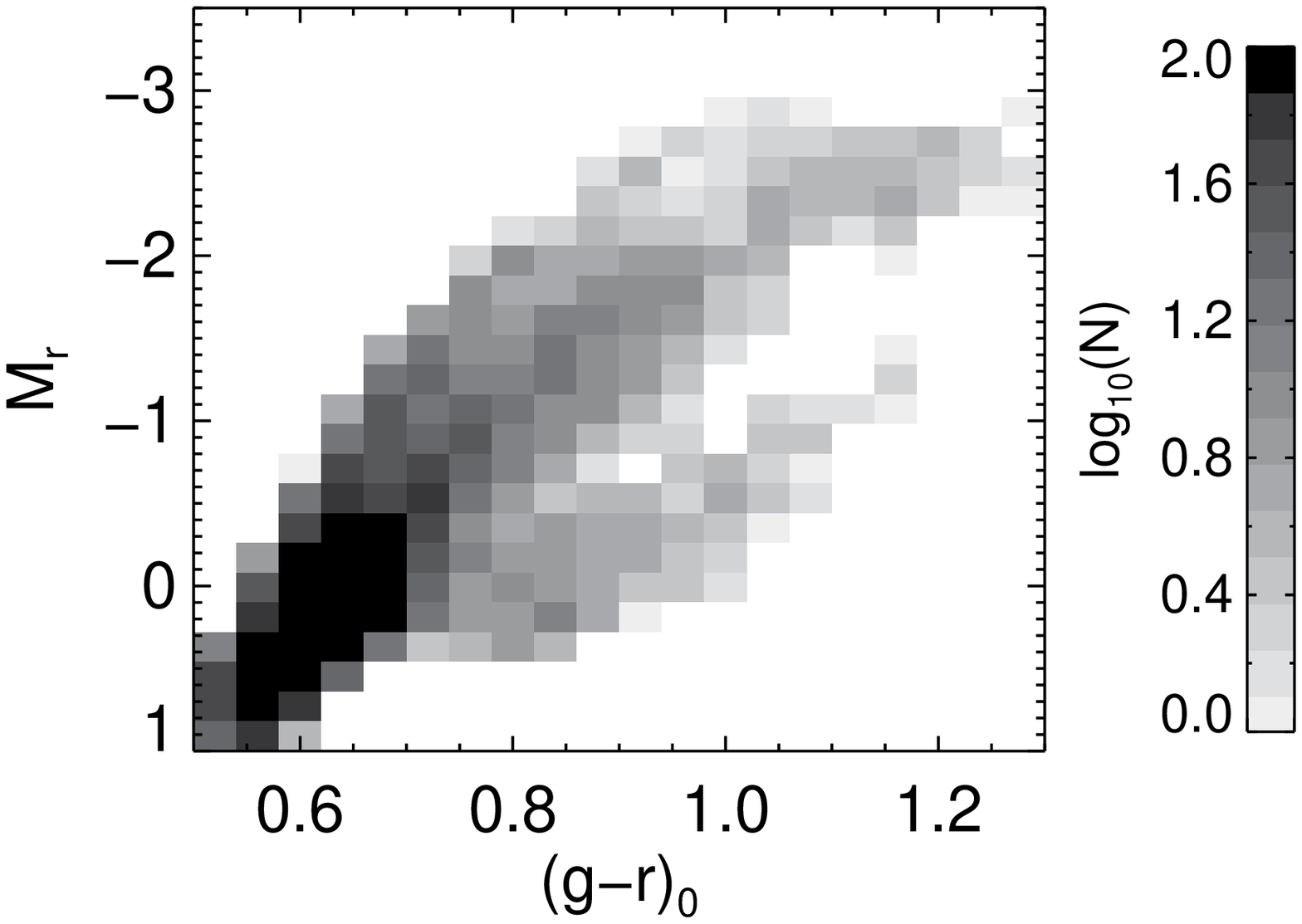}
\includegraphics[width=0.88\textwidth,height=0.45\textheight]{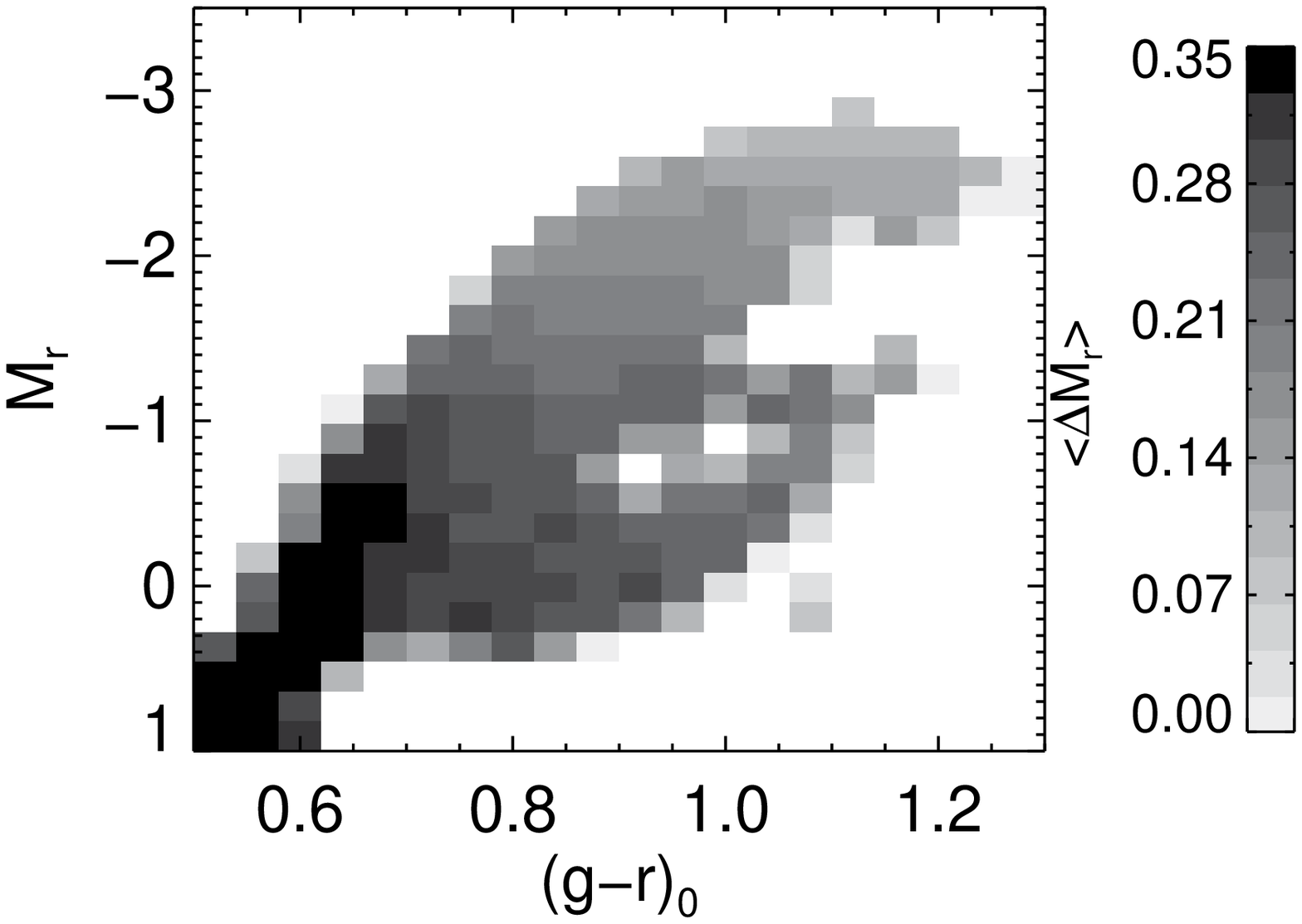}
\caption{(Upper panel) The distribution of K giants on the CMD
  plot. (Lower panel) The distribution of the mean error in the
  absolute magnitude, $\rm M_{\it r}$, as a function of $\rm M_{\it r}$
  and $(g-r)_0$. The upper panel shows that the sample contains a
  large fraction of relatively nearby giants of moderate luminosity
  ($\rm M_{\it r} \sim 0$). (Lower panel) The luminosity estimates for stars in the
  lower part of the CMD are less precise, because the
  isochrones are much steeper in this part, especially for low
  metallicities.}
\label{f:fcmd}
\end{figure}

The bottom panel of Figure~\ref{f:fresult} shows that the mean error
in $\rm M_{\it r}$ is $\rm \pm 0.35$ mag, and that faint giants have
less precise intrinsic luminosity estimates. Figure~\ref{f:fcmd}
(upper panel) shows the distribution of K giants on the CMD. There are
more stars in the lower part of the giant branch, which is consistent
with the prediction of the luminosity function. The lower panel of
Figure~\ref{f:fcmd} shows that stars in the upper part of the RGB have
more precise distances than those in the lower part of the CMD,
because the fiducials are much steeper near the sub-giant branch. This
is equivalent to the fact that the fractional distance precision is
higher for the largest distances (see Figure~\ref{f:fresult}).

The giants in our sample lie in the region of $\rm 5-125$ kpc
from the Galactic center. There are 1647 stars
beyond 30 kpc, 283 stars beyond 50 kpc, and 43 stars beyond 80 kpc
(c.f., the $\rm 5$ red giants beyond $\rm 50$ kpc in
\citealt{Battaglia2005}, $\rm 16$ halo stars beyond $\rm 80$ kpc in
\citealt{Deason2012} and no BHB stars beyond $\rm 80$ kpc in
\citealt{Xue2008,Xue2011}). Our sample comprises the largest sample of
distant stellar halo stars with measured radial velocities and
distances to date.

\subsection{The Impact of Priors}
In this section we briefly analyze how important the priors actually
were in deriving the distance estimates. For each star, the evaluation
of Eq. 3 using Eq. 4 and the interpolated fiducials results in $\rm
\ldm$ (Figure~\ref{f:fldm}), i.e., the likelihood of the distance
modulus, {\it before} folding in an explicit prior on $\rm \DM$, but
{\it after} accounting for the priors on M and $\rm \feh$ (Eq. 3). In
this section we present some example $\rm \ldm$, but first explore the systematic impact on $\rm \DM$ of neglecting the $\rm M$
and $\rm \feh$ priors.

When estimating the distance to a given star, without the benefit of
external prior information, one would evaluate Eq. 3 presuming that
$\rm p_{prior}\bigl(M\bigr)$ and $\rm p_{prior}\bigl(\feh\bigr)$ are
constant.

\begin{figure}
\includegraphics[width=\textwidth,height=0.8\textheight]{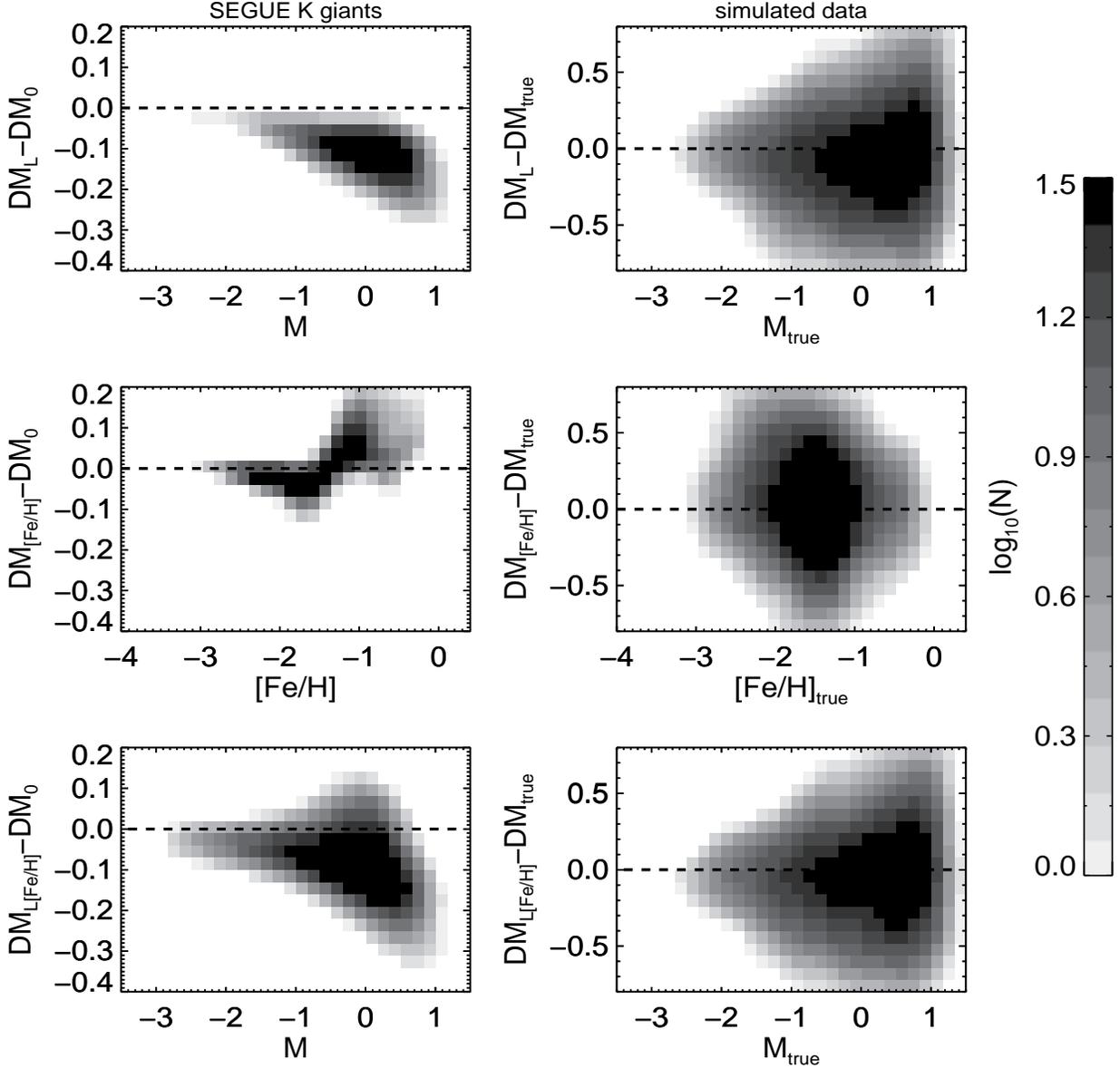}

\caption{The left panel shows the distance modulus bias caused by neglecting the priors on
  the luminosity function and metallicity distribution of the K
  giants. The luminosity-function prior can help correct a mean
  overestimate of $\rm 0.1$ mag in the distance moduli, and a maximum
  overestimate of $\rm \sim 0.25$ mag in some cases. While the impact
  of $\rm \feh$ prior is smaller, it can help correct a mean of
  $\rm 0.03$ mag overestimate, or a mean of $\rm 0.05$ mag underestimate on the distances in the
  metal-poor or metal-rich tails. The bottom panel shows the total
  impacts of the luminosity prior and the metallicity prior. The right panel shows the comparison between the true distance modulus and the calculated ones for the simulated data. For the cases from the top to the bottom, the mean values and sigmas of $\bigl(\rm
  \DM_{cal.}-\DM_{true}\bigr)/\rm \sigma_{\DM_{cal.}}$ are (-0.14,0.95),(0.17,0.95) and (-0.09,0.94), which shows including both priors can lead to the most consistent distance modulus.}
\label{f:fbias}
\end{figure}

To test the impact of $\rm p_{prior}\bigl(M\bigr)$, we estimate the
distances for two cases. No priors applied ($\rm
p_{prior}\bigl(M\bigr)=1$ and $\rm p_{prior}\bigl(\feh\bigr)=1$); and
only the prior on the luminosity function applied ($\rm p\bigl(L\bigr)
\propto L^{-1.8}$ and $\rm p_{prior}\bigl(\feh\bigr)=1$). The distance
modulus estimate that neglects the explicit priors is denoted as $\rm
DM_0$, while the distance modulus with only $\rm
p_{prior}\bigl(M\bigr)$ applied is marked as $\rm DM_L$. The top of the left panel of Figure~\ref{f:fbias} illustrates the importance of including the `luminosity prior', by showing the systematic difference in $\rm \DM$ that results from neglecting it. For stars near the tip of the giant branch the bias is very small, but for stars near the bottom of the
giant branch, the mean systematic bias of neglecting the luminosity prior information is $\rm 0.1$ mag, with systematic bias as high as $\rm \sim 0.25$ mag in some cases.

To test the impact of $\rm p_{prior}\bigl(\feh\bigr)$, we estimate the
distances where only the metallicity prior was applied, and mark the
relevant $\rm \DM$ as $\rm DM_{\feh}$. Compared with the distance
modulus with no priors applied, $\rm DM_0$, we find $\rm
p_{prior}\bigl(\feh\bigr)$ can correct a mean overestimate of $\rm
0.03$ mag on the $\rm \DM$ for the metal-poor stars and a mean
underestimate of $\rm 0.05$ mag on the $\rm \DM$ for the metal-rich
ones, but the neglect of the metallicity prior causes a smaller bias
in $\rm \DM$ than neglecting the luminosity prior ($\rm 0.05~mag ~{\it
  vs.}~ 0.1$ mag at mean), as shown in Figure~\ref{f:fbias} (middle
of left panel). The distance modulus bias caused by neglecting both priors is
presented in the bottom panel of Figure~\ref{f:fbias}. Neglecting both
priors causes a mean bias of $\rm 0.1$ mag and a maximum bias of $\rm
0.3$ mag in distance modulus.

\subsection{Distance Precision Test using Clusters}

We use five clusters (M13, M71, M92, NGC6791, and Berkeley29)
  to test the precision of the distance estimates, because they have
  spectroscopic members observed in SEGUE. Berkeley29 (Be29) is a comparatively young open cluster with age of 3$\sim$4 Gyr \citep{Sestito2008}, younger than our adopted fiducials (10$\sim$12
  Gyr). It illustrates how distances could be in error as a result of an incorrect age prior. M71 is
  a disk globular cluster. Because of its low Galactic latitude (less than 5 degrees from the disk plane) and relatively circular orbit, separation of genuine M71 members from field stars is much more difficult than for halo clusters (M13 and
  M92). The analysis concerning the membership of stars in the stellar clusters will be reported in detail in the Appendix of the Morrison et al. (in prep.).

In general, we identify cluster membership using proper motion and radial
  velocity. The proper motions provide a membership probability for each star\citep{Cudworth1976,Cudworth1979,Cudworth1985,Rees1992}, and then the radial velocities are used for further membership checks, as described in detail in Morrison et al. ({\it in
    preparation}).

\begin{figure}
\includegraphics[width=\textwidth]{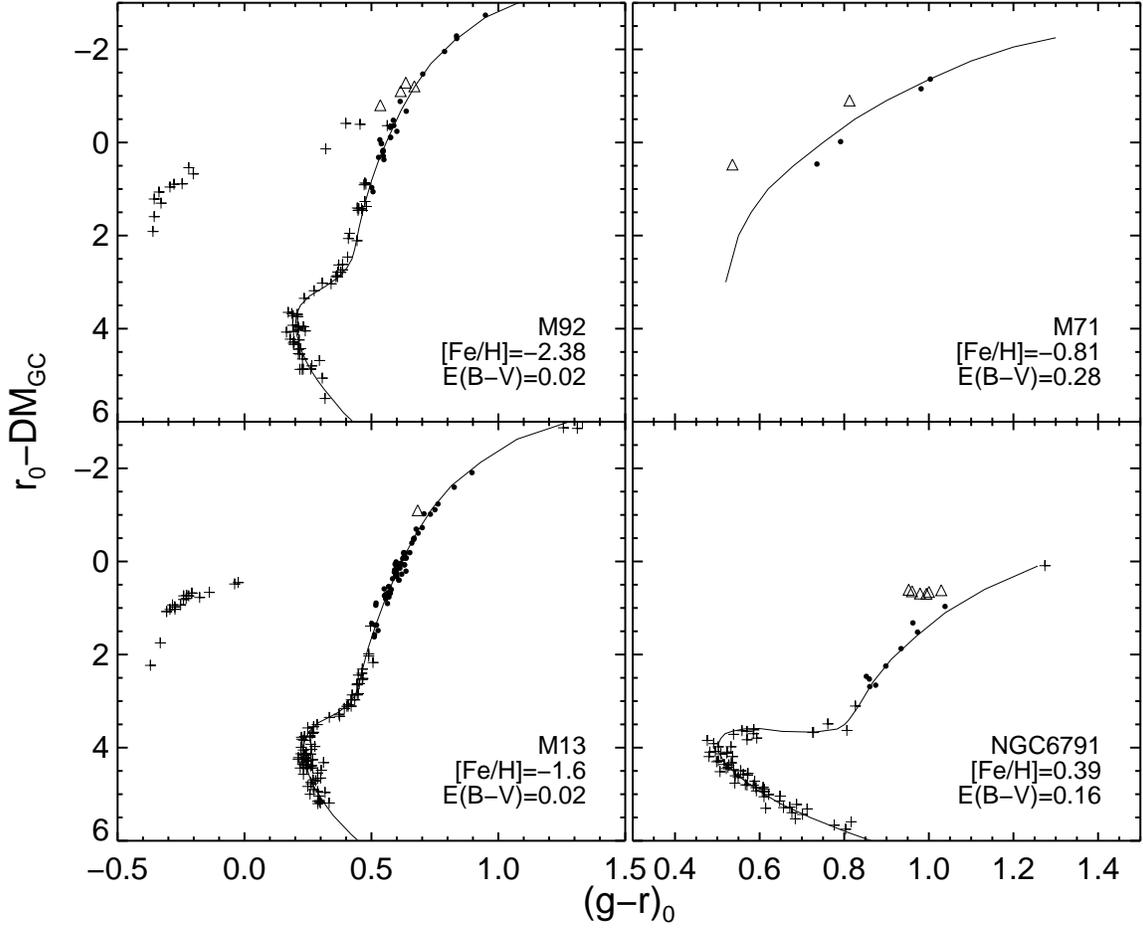}
\caption{The color-magnitude diagrams for the four clusters used both
  for fiducials and distance precision test. The solid lines are the
  fiducials derived by the photometry. Only member stars observed in
  SEGUE are over-plotted. The filled circles are RGB member stars used to test the distance precision, the triangles are non-RGB stars (i.e., HB/RC or AGB stars), and the plus signs are main-sequence stars or RGB stars with $|\feh_{member}-\feh_{GC}|>0.23$ dex.}
\label{f:fcmdcluster}
\end{figure}

Based on the color-magnitude diagram (CMD) of the clusters, we select giant members with $(g-r)_0>0.5$ and above the sub-giant branch of the clusters to test the distance precision. The range of signal-to-noise ratio for the spectra of the cluster giant stars is $[10,70]$, and the SN rang for K-giant spectra is $[10,120]$. Here we do not apply the very stringent criterion to eliminate HB/RC stars described in \S 3.4, because this criterion also culls many lower RGB stars that are useful for the test. Figure~\ref{f:fcmdcluster} shows there are some HB/RC stars or AGB stars, which can help verify how distances would be in error for the non-RGB stars. Furthermore, we do not use the members with $\rm |\feh_{member}-\feh_{GC}|>0.23$ dex, because of the strong sensitivity of the distances to metallicity errors.

\begin{figure}
\includegraphics[width=\textwidth,height=\textwidth]{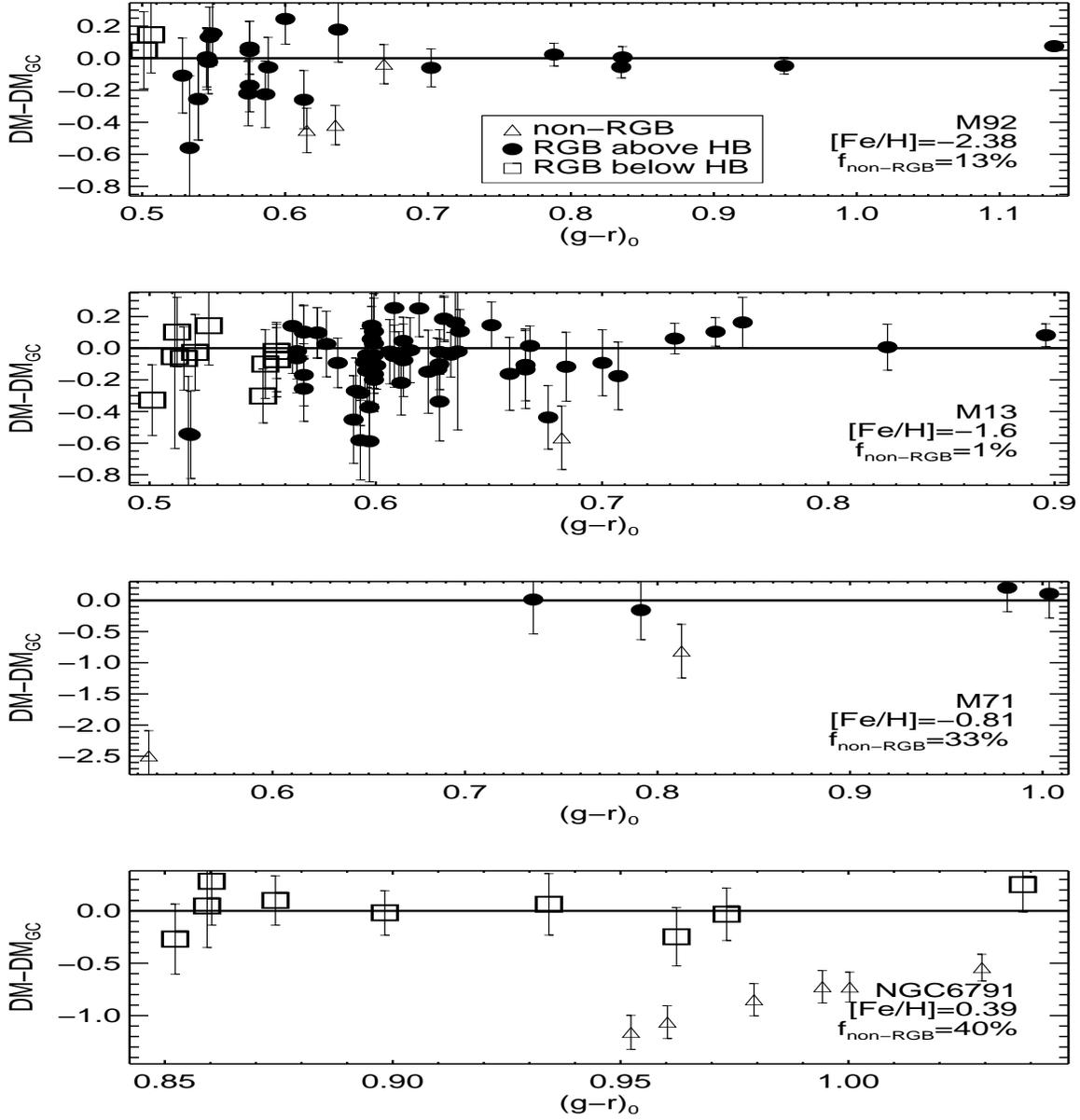}
\caption{The differences between individual $\rm \DM$ estimated by our Bayesian approach for RGB and non-RGB member stars and the literature $\rm \DM_{GC}$. The filled circles and squares are both RGB members, lying above or below the horizontal-branch, respectively,  according to the relation between $(g-r)^{HB}_0$ and $\rm \feh$ in \S 3.4, which shows that the recovered values of $\rm \DM$ are consistent with the literature $\rm \DM_{GC}$ within 1$-\sigma$. The triangles are non-RGB stars (i.e., HB/RC or AGB stars), for which the distance estimates are underestimated by up to $1.24$ mag. This shows the criterion to eliminate HB/RC stars adopted in \S 3.4 is sufficiently stringent to cull all HB/RC stars and many lower-RGB stars.}
\label{f:fcluster}
\end{figure}

We estimate the $\rm \DM$ for each selected member RGB star, adopting the luminosity prior of $\rm
  p(L) \propto L^{-1.8}$, and a Dirac delta function centered at the literature cluster metallicity, $\rm \feh_{GC}$, as the metallicity
  prior. Figure~\ref{f:fcluster} shows the difference between our individual $\rm \DM$ estimate for each selected member RGB star and the literature $\rm \DM_{GC}$ (shown in Table 1) for M13, M71, M92 and NGC6791 respectively. Since M71 has significant differential reddening and fewer members, it is not a suitable cluster to verify the distance errors to little-reddened, usually more metal-poor halo giants. However, all 4 M71 members exhibit less than $\rm 0.2$ mag scatter from $\rm \DM_{GC}$, as shown in Figure~\ref{f:fcluster}. The RGB members show consistent distances with the literature value derived by main-sequence fitting within 1$-\sigma$, but the distance moduli are underestimated by a maximum $\rm 1.24$ mag for the non-RGB stars. Fortunately, the criterion to eliminate HB/RC stars adopted in \S 3.4 is sufficiently stringent to cull all HB/RC stars and many lower-RGB stars.  As mentioned previously, there is a relative paucity of AGB stars in the SEGUE K-giant sample.

The mean values of $\rm (\DM-\DM_{GC})$ for the RGB members at different color ranges are within $\pm 0.1$ mag. Compared to the typical error of $0.35$ mag in $\rm \DM$, our estimates of $\rm \DM$ are reasonably precise.

\begin{figure}
\includegraphics[width=\textwidth]{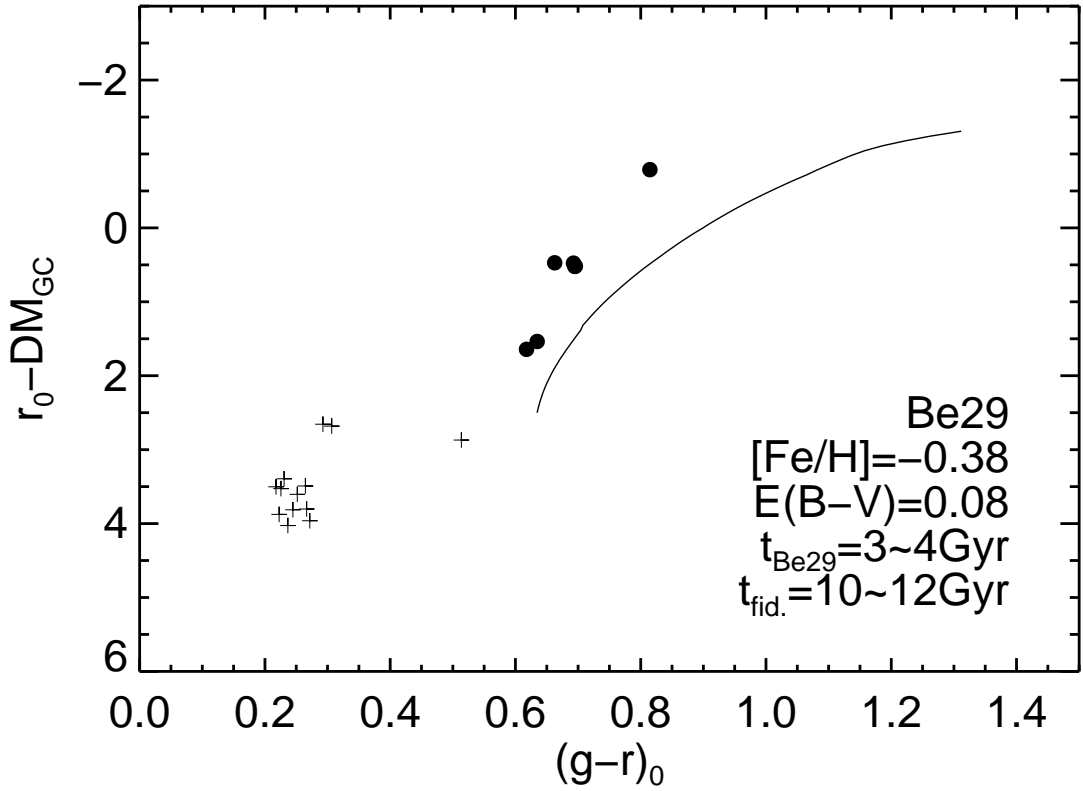}
\caption{The color-magnitude diagram for Be29. The solid lines are
  the interpolated fiducial based on Figure~\ref{f:ffiducial}. The plus signs and filled circles are member stars of the cluster observed in SEGUE, while the filled
  circles are the RGB members. The interpolated fiducial is older than the cluster, so it leads to incorrect distance estimates, as shown in Figure~\ref{f:fcluster2}.}
\label{f:fcmdcluster2}
\end{figure}

\begin{figure}
\includegraphics[width=\textwidth]{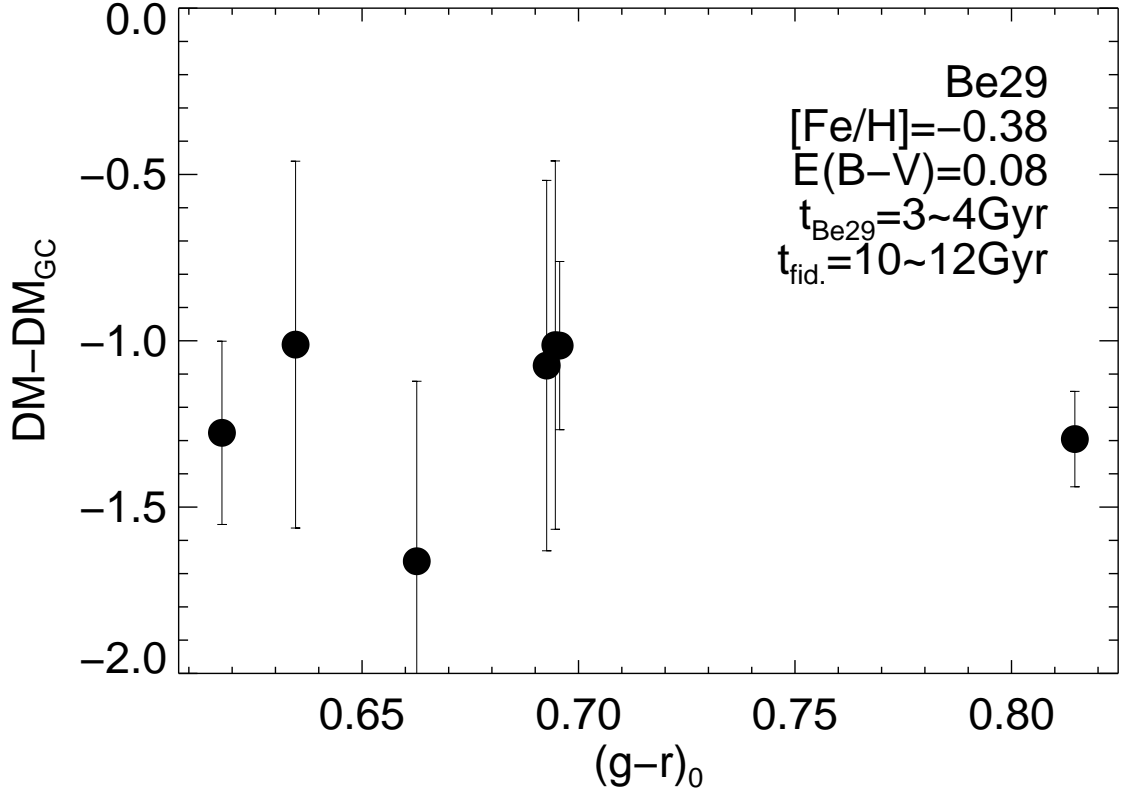}
\caption{The differences between individial $\rm \DM$ for spectroscopic RGB members and the literature $\rm \DM_{GC}$ for Be29. Filled circles are the RGB member stars, which shows that the distance estimates based on the old fiducials are incorrect, due to the use of the wrong age prior.}
\label{f:fcluster2}
\end{figure}

Figure~\ref{f:fcmdcluster2} shows the distributions of Be29 members around their fiducials. Be29 members are far from the interpolated fiducial based on our old fiducials. The old fiducials lead to totally wrong distance estimates for relatively young giants in Be29, as shown in Figure~\ref{f:fcluster2}. If the age prior is wrong, the distance estimates are unreliable. The derived errors on the distance moduli of the K giants are only valid if the ages of the K giants are older than 10 Gyr. 

In addition, we calculate the distances with flat priors (which means no priors), and find that neglecting the priors would lead to biases in distance of $(6\%,6\%,3\%,0.7\%)$ from the literature values for NGC6791, M71, M13 and M92. However, we only find biases in distance of $(0.7\%,2\%,2\%,0.4\%)$ the literature values for NGC6791, M71, M13 and M92 when using both priors. Therefore, neglecting the priors would lead to biased distance estimates. 


\section{Summary and Conclusions}
We have implemented a probabilistic approach to estimate the distances
for SEGUE K giants in the Galactic halo. This approach folds all
available observational information into the calculation, and
incorporates external information through priors, resulting in a $\DM$
likelihood for each star that provides both a distance estimate and its
uncertainty.

The priors adopted in this work are the giant-branch luminosity
function derived from globular clusters, and the ensemble metallicity
distributions for different SEGUE K-giant target categories. We show
that these priors are needed to prevent systematic overestimate of the
distance moduli by up to $\rm 0.25$ mag. The role of the priors are important,
and make the results more reliable.

We employed empirical color-magnitude fiducials from old stellar clusters
to obtain the predicted colors $\rm {\it c}\bigl(M,\feh\bigr)$, which are
needed to calculate $\rm \ldm$. Ultimately, the best estimates of the
distance moduli and their errors can be estimated using the peak and
central $\rm 68\%$ interval of $\rm \ldm$.

We used a simulated data set to verify that our distance estimates are close to optimal and are nearly unbiased: any systematic biases in distance will be of
  order 1\%. We verified that the exact form of the prior for the halo density profile
  does not affect our results significantly, which is consistent with
  the conclusion of \citet{burnett2010}. Neglecting the luminosity
  and metallicity priors, as has often been done in previous distance analyses, will lead to a mean systematic bias in distance modulus of $\rm 0.1$ mag, which shows the advantage of our Bayesian method.

With this approach we obtain the distance moduli,
and thus, absolute magnitudes and distances, for $\rm 6036$ K giants
that have a mean distance precision of $\rm 16\%$, or $\rm \pm 0.35$
mag in $\rm \DM$ and $\rm M_{\it r}$. The sample contains $\rm 283$
stars beyond $\rm r_{GC}=50$ kpc, which makes it by far the largest
sample of distant stellar halo stars with measured radial velocities
and distances to date.

We tested the accuracy of our distance estimates using RGB member stars
  in the clusters M13, M92, M71, NGC6791 and Be29, which have
  SEGUE spectroscopic observations. We found that the distance
  estimates for the individual cluster member RGB star derived with Bayesian approach are consistent with the literature distance moduli of the clusters within 1$-\sigma$. In addition, we found our fiducials would lead to substantially incorrect distance estimates for young giants ($\rm <5$ Gyrs), based on our test for Be29.

We present an online catalog containing the distance moduli, observed
information and SSPP atmospheric parameters for the 6036 SEGUE K
giants. For each object in the catalog we also list some of the basic
observables such as (RA, Dec), extinction-corrected apparent magnitudes
and de-reddened colors, as well as the information obtained from the spectra--
heliocentric radial velocities plus SSPP atmospheric parameters. In
addition, we provide the Bayesian estimates of the distance moduli,
distances to the Sun, Galactocentric distances, the absolute
magnitudes and their uncertainties, along with the distance moduli at
$\rm (5\%,16\%, 50\%,84\%,95\%)$ confidence of $\rm \ldm$.

We caution the reader that the $\rm \rm n(d,$\rm \feh$)$ cannot be
used to obtain the halo profile and the metallicity distribution
directly, because the complex SEGUE selection function needs to be
taken into account.

\acknowledgments

Funding for SDSS-III has been provided by the Alfred P. Sloan
Foundation, the Participating Institutions, the National Science
Foundation, and the U.S. Department of Energy Office of Science. The
SDSS-III web site is http://www.sdss3.org/.

SDSS-III is managed by the Astrophysical Research Consortium for the
Participating Institutions of the SDSS-III Collaboration including the
University of Arizona, the Brazilian Participation Group, Brookhaven
National Laboratory, University of Cambridge, Carnegie Mellon
University, University of Florida, the French Participation Group, the
German Participation Group, Harvard University, the Instituto de
Astrofisica de Canarias, the Michigan State/Notre Dame/JINA
Participation Group, Johns Hopkins University, Lawrence Berkeley
National Laboratory, Max Planck Institute for Astrophysics, Max Planck
Institute for Extraterrestrial Physics, New Mexico State University,
New York University, Ohio State University, Pennsylvania State
University, University of Portsmouth, Princeton University, the
Spanish Participation Group, University of Tokyo, University of Utah,
Vanderbilt University, University of Virginia, University of
Washington, and Yale University.

This work was made possible by the support of the Max-Planck-Institute
for Astronomy, and supported by the National Natural Science
Foundation of China under grant Nos. 11103031, 11233004, 11390371 and 11003017,
and supported by the Young Researcher Grant of National Astronomical
Observatories, Chinese Academy of Sciences. This paper was partially
supported by the DFG's SFB-881 grant `The Milky Way System'.

X.-X. Xue acknowledges the Alexandra Von Humboldt foundation for a fellowship

H. Morrison acknowledges funding of this work from NSF grant
AST-0098435.

YSL and TCB acknowledge partial support of this work from grants PHY
02-16783 and PHY 08-22648: Physics Frontier Center / Joint Institute
for Nuclear Astrophysics (JINA), awarded by the U.S. National Science
Foundation.

HRJ acknowledges support from the National Science Foundation under
award number AST-0901919.

JJ acknowledges NSF's grants AST-0807997 and AST-0707948.

SL reasearch is partially supported by the INAF PRIN grant "Multiple
populations in Globular Clusters: their role in the Galaxy assembly".
\bibliographystyle{apj}
\bibliography{refkg}
\clearpage

\begin{table}
\label{cluster}
\centering
\begin{threeparttable}
\caption{Parameters of the Fiducial Clusters and Be29}
\newcolumntype{R}{>{\raggedleft\arraybackslash}X}
\begin{tabularx}{0.7\textwidth}{RRRRR}
\hline\hline\\ NGC& Messier& E(B-V)& $(m-M)_0$ & $\rm \feh$
\\ \hline\\ 6341&M92&0.02\tablenotemark{a} &14.64\tablenotemark{b} &
--2.38\tablenotemark{a} \\ 6205&M13&0.02\tablenotemark{a}
&14.38\tablenotemark{b} &--1.60\tablenotemark{a}
\\ 6838&M71&0.28\tablenotemark{c} &12.86\tablenotemark{c} &
--0.81\tablenotemark{a} \\ 6791& &0.16\tablenotemark{d}
&13.01\tablenotemark{d} &+0.39\tablenotemark{e}\\ &
Be29\tablenotemark{f}&0.08&15.6&--0.38\\ \hline
\end{tabularx}

\begin{tablenotes}
\item 

\tablenotetext{a}{\citet{kra03}; their globular cluster metallicity
  scale is based on the Fe$\rm II$ lines from high-resolution spectra
  of giants.}
  
\tablenotetext{b}{\citet{car00}; $(m-M)_0$ derived from the Hipparcos
  sub-dwarf fitting.}
  
\tablenotetext{c}{\citet{gru02}; $(m-M)_0$ derived from the Hipparcos
  \citep{per97} sub-dwarf fitting.}
  
\tablenotetext{d}{\citet{bro11}; $(m-M)_0$ is based on $(m-M)_v$
  assuming $A_v=3.1E(B-V)$.}  \tablenotetext{e}{simple average of the
  $\rm \feh$ +0.29, +0.47, +0.4 and +0.39 by \citet{bro11},
  \citet{gra06}, \citet{pet98}, and \citet{car06} respectively.}

\tablenotetext{f}{reddening is from \citet{Carraro2004}, $(m-M)_0$ is
  from \citet{Sestito2008} and $\rm \feh$ is the average of the values
  from \citet{Carraro2004} and \citet{Sestito2008}.}

\end{tablenotes}
\end{threeparttable}
\end{table}

\begin{table}
\label{hb}
\centering
\begin{threeparttable}
\caption{Metallicity and Color of the Red Horizontal-Branch Onset for
  the Eight Clusters in \citet{An2008}}
\newcolumntype{R}{>{\raggedleft\arraybackslash}X}
\begin{tabularx}{0.6\textwidth}{cRR}
\hline\hline\\ $\rm Name ~of~ Clusters $ & $\rm \feh$ & $(g-r)_0^{HB}$
\\ \hline\\ 

NGC6791&+0.39&1.13\\ 
M71&--0.81&0.69\\
M5&--1.26&0.61\\ 
M3&--1.50&0.59\\ 
M13&--1.60&0.58\\ 
M53&--1.99&0.54\\ 
M92&--2.38&0.53\\ 
M15&--2.42&0.53\\ 
\hline

\end{tabularx}

\begin{tablenotes}
\item The first column lists the names of the clusters; the next two
  columns provide the $\rm \feh$ of the clusters and the extinction
  corrected color $(g-r)_0$.

\end{tablenotes}
\end{threeparttable}
\end{table}

\begin{table}
\label{cmd}
\centering
\begin{threeparttable}
\caption{Interpolated Fiducial at $\rm \feh$=--1.18}
\newcolumntype{R}{>{\raggedleft\arraybackslash}X}
\begin{tabularx}{0.4\textwidth}{cRR}
\hline\hline\\
$\rm (g-r)_0 $   & $\rm M_{\it r}$\\
\hline\\
0.473&   3.000\\ 
0.491&   2.546\\ 
0.527&  1.747\\  
0.545&   1.470\\ 
0.581&   0.985\\ 
0.599&   0.790\\ 
0.635&   0.429\\ 
0.671&   0.087\\ 
0.706&  -0.237\\ 
0.742&  -0.523\\ 
0.778&  -0.782\\ 
0.814&  -1.013\\ 
0.850&  -1.225\\ 
0.886&  -1.422\\ 
0.921&  -1.596\\ 
0.957&  -1.746\\ 
0.993&  -1.888\\ 
1.029&  -2.010\\ 
1.065&  -2.116\\ 
1.101&  -2.227\\ 
1.136&  -2.321\\ 
1.172&  -2.403\\ 
1.208&  -2.477\\ 
1.244&  -2.545\\ 
1.280&  -2.609\\ 
1.316&  -2.668\\ 
1.351&  -2.725\\ 
\hline

\end{tabularx}

\begin{tablenotes}
\item An example of one interpolated fiducial with $\rm
  \feh$=--1.18. The entire catalog of 50 interpolated fiducials with metallicity ranging from $\rm \feh=-2.38$ to $\rm \feh=+0.39$ is provided in the electronic edition of the journal.

\end{tablenotes}
\end{threeparttable}
\end{table}

\begin{sidewaystable}
\label{sample}
\begin{threeparttable}
\tiny
\caption{List of $\rm 6036$  K Giants Selected from SDSS DR9}
\newcolumntype{R}{>{\raggedleft\arraybackslash}X}
\begin{tabularx}{\textwidth}{ccRRRRRRRRRRR}
\hline\hline\\ 
$\rm RA (J2000)$ & $\rm Dec (J2000)$ & $r_0$ & $\Delta
r_0$ & $(g-r)_0$ & $\Delta (g-r)_0$ & $\rm RV$ & $\rm \Delta RV$ &
$\rm T_{eff}$ & $\rm \feh$ & $\rm \Delta \feh$&$\rm log~g$ & $\rm \DM_{peak}$ \\ 
$\rm (deg)$ & $\rm (deg)$ & $\rm (mag)$ & $\rm (mag)$ &
$\rm (mag)$ & $\rm (mag)$ & (km$s^{-1}$) & (km$s^{-1}$) & (K) & 
&  & & $\rm (mag)$\\ 
\hline\\ 
154.7659&        -0.8354&    17.257&     0.040&     0.965&     0.028&    43.7&     2.5&    4626&   -0.71&    0.16&    3.28&   18.13\\
174.6570&        -0.9330&    17.036&     0.041&     0.590&     0.035&   147.0&     3.6&    5259&   -1.50&    0.15&    2.59&   16.37\\
189.9634&         1.0202&    17.078&     0.040&     0.558&     0.028&   122.7&     6.1&    5296&   -1.89&    0.19&    2.99&   16.41\\
196.0723&        -0.5404&    16.891&     0.041&     0.526&     0.030&  -169.5&     4.2&    5158&   -2.12&    0.15&    1.75&   16.03\\
205.9106&        -0.3442&    18.067&     0.041&     0.631&     0.031&    86.5&     4.5&    4901&   -1.23&    0.21&    1.98&   17.68\\       
\hline\\ \hline\hline\\

$\rm \DM_{5\%}$& $\rm \DM_{16\%}$ & $\rm \DM_{50\%}$ & $\rm
\DM_{84\%}$ & $\rm \DM_{95\%}$ & $\rm \Delta \DM$ & $\rm M_{\it r}$ &
$\Delta M_{\it r}$ & $\rm d$ & $\Delta d$ & $\rm r_{GC}$ & $\rm \Delta
r_{GC}$&$\rm P_{aboveHB}$\\ $\rm (mag)$ & $\rm (mag)$ & $\rm (mag)$ & $\rm (mag)$ &
$\rm (mag)$ & $ \rm (mag)$ & (mag) &(mag) & $\rm (kpc)$ & $ \rm (kpc)$
& (kpc)&(kpc)&\\

\hline\\ 
 17.62&   17.85&   18.21&   18.62&   18.83&    0.38&   -0.88&    0.39&   42.35&    7.89&   45.51&    7.78&    1.00\\   
15.43&   15.81&   16.33&   16.79&   17.07&    0.49&    0.67&    0.49&   18.78&    4.11&   20.55&    3.79&     0.67\\
15.56&   15.90&   16.39&   16.86&   17.14&    0.48&    0.66&    0.48&   19.17&    4.18&   19.27&    3.82&     0.75\\
15.00&   15.40&   15.96&   16.46&   16.76&    0.53&    0.86&    0.53&   16.05&    3.79&   15.64&    3.31&     0.51\\
17.00&   17.29&   17.68&   18.07&   18.32&    0.39&    0.39&    0.39&   34.38&    6.24&   31.74&    6.07&     0.66\\
\hline
\end{tabularx}

\begin{tablenotes}
\item The first two columns list the position (RA, Dec) for each
  object. The magnitudes,colors, and their errors are provided in the next four
  columns: corrected for extinction. The heliocentric radial
  velocities and their errors are listed in the next two columns. The
  next four columns contain the stellar atmospheric parameters and the
  errors in the metallicities as a relation of signal-to-noise
  ratio. Effective temperatures and surface gravities are not used in
  our work, and they are all published in SDSS DR9, so we recommend
  interested readers download their errors from CasJob. The $\rm
  \DM$ at the peak and $\rm (5\%,16\%, 50\%,84\%,95\%)$ confidence of
  $\rm \ldm$ are listed in the next six columns. $\rm \DM_{peak}$ is
  the best estimate of the distance modulus for the K giant. The $\rm
  \Delta_{\DM}$ is the uncertainty of the distance modulus, which is
  calculated from $\rm (\DM_{84\%}-\DM_{16\%})/2$. The last seven
  columns are absolute magnitude and distances calculated from $\rm
  \DM_{peak}$, assuming $\rm R_\odot=8.0$ kpc (i.e., $\rm M_{\it
    r}={\it r_0}-\DM_{peak}$, $\rm d=10^{\frac{\DM+5}{5}}$), and the chance of being clearly RGB. The
  complete version of this table is provided in the electronic edition
  of the journal. The printed edition contains only a sample.

\end{tablenotes}
\end{threeparttable}
\end{sidewaystable}

\end{document}